\documentclass[aps,preprintnumbers,twocolumn,amsmath,amssymb,floatfix,superscriptaddress,pra]{revtex4-1}
\pdfoutput=1
\usepackage[utf8]{inputenc}
\usepackage{amsmath}
\usepackage{amsfonts}
\usepackage{braket}
\usepackage{tensor}
\usepackage{graphicx}
\usepackage[colorlinks=true,linkcolor=blue, citecolor=blue, urlcolor=blue, bookmarks]{hyperref}

\def\be{\begin{equation}}
\def\ee{\end{equation}}
\def\bea{\begin{eqnarray}}
\def\eea{\end{eqnarray}}

\def\XXint#1#2#3{{\setbox0=\hbox{$#1{#2#3}{\int}$}
         \vcenter{\hbox{$#2#3$}}\kern-.5\wd0}}

\begin{document}

\title{Spin-charge separation effects in the low-temperature transport of 1D Fermi gases}
\author{M\'arton Mesty\'an}
\affiliation{SISSA and INFN, via Bonomea 265, 34136 Trieste, Italy.}
\author{Bruno Bertini}
\affiliation{Department of physics, FMF, University of Ljubljana, Jadranska 19, SI-1000 Ljubljana, Slovenia}
\author{Lorenzo Piroli}
\affiliation{SISSA and INFN, via Bonomea 265, 34136 Trieste, Italy.}
\author{Pasquale Calabrese}
\affiliation{SISSA and INFN, via Bonomea 265, 34136 Trieste, Italy.}
\address{International Centre for Theoretical Physics (ICTP), I-34151, Trieste, Italy.}

\begin{abstract}
We study the transport properties of a one-dimensional spinful Fermi gas, after junction of two semi-infinite sub-systems held at different temperatures. The ensuing dynamics is studied by analysing the space-time profiles of local observables emerging at large distances $x$ and times $t$, as a function of $\zeta=x/t$. At equilibrium, the system displays two distinct species of quasi-particles, naturally associated with different physical degrees of freedom. By employing the generalised hydrodynamic approach, we show that when the temperatures are finite no notion of separation can be attributed to the quasi-particles. In this case the profiles can not be qualitatively distinguished by those associated to quasi-particles of a single species that can form bound states. On the contrary, signatures of separation emerge in the low-temperature regime, where two distinct characteristic velocities appear. In this regime, we analytically show that the profiles display a piece-wise constant form and can be understood in terms of two decoupled Luttinger liquids. 
\end{abstract}

\maketitle

\section{Introduction}\label{sec:intro}
Hydrodynamic approaches provide the skeleton of our understanding of many-body quantum physics. This is especially true in generic situations, when a full microscopic description is out of reach. In particular, quantum hydrodynamics has proved to be extremely powerful in one dimension, where seminal works by Haldane \cite{Hald81} have set the basis for the development of the well-established theory of Luttinger liquids \cite{giamarchi-04,GNTbook}.This approach consists in describing the excitations over the ground state of the system as a gas of non-interacting particles with linear dispersion. Quantitative predictions can be obtained for large-distance correlations \cite{Caza04}, which are fully determined by the low-momentum and low-energy modes. Further progress was made by also taking into account effects of non-linearity~\cite{nonlinearhydro}, resulting, for instance, in the predictions of universal properties beyond the linear approximation \cite{ImGl09,ImGl09-science,ImScGl-review,KaPS15}. 

Even though information on the microscopic, short-scale structure of the system is lost, the Luttinger theory has proven to be an excellent tool to quantitatively describe many genuinely quantum effects~\cite{giamarchi-04}. A prominent example is given by the well-known phenomenon of spin-charge separation \cite{giamarchi-04,GNTbook,EFGK_book,KMSM96,SPHB99,ASYT05,JFGF09,RFZZ03,KoSZ05,PoVi07,KKMG08,ScIG10,ScIG10-2}. In a 
one-dimensional system of interacting quantum particles with spin, the spin and charge of its constituents might effectively split and propagate separately with different velocities. The explanation of this effect lies in the collective nature of the elementary excitations, which can not be understood uniquely in terms of the physical degrees of freedom.

Recently, a different hydrodynamic approach has been introduced in the context of quantum integrable systems \cite{CaDY16,BCDF16}. Within this framework, often called generalised hydrodynamics, one is still interested in a large-scale description of the system, but its scope is not restricted to the low-energy sector. This approach provides \emph{exact} predictions in the limit of infinite distance-and time-scales~\cite{CaDY16,BCDF16,PDCB17}, but it also gives extremely good approximations for large but finite scales \cite{BVKM17,BVKM17-2,DDKY17,DoYC17,Fago17}, and it has successfully been employed to study spreading of entanglement and correlations~\cite{DoCorr17,IlDe17,cdv-17,MVCR18, Alba17, BFPC18, IDMP18,a-18b}. The crucial ingredient of this approach is the existence of stable quasi-particles: this makes it possible to adopt a description in terms of space- and time-dependent local quasi-stationary states \cite{BeFa16}, which are fully determined by their quasi-momenta distribution functions. The evolution of such distributions is obtained by following the motion of the quasi-particles and results in a set differential (or continuity) equations.

Analogously to the traditional (quantum) hydrodynamics, this approach is able capture several hallmarks of collective quantum phenomena. To analyse these phenomena is often convenient to consider minimal settings where a detailed study can be carried out. The most popular consists of taking two semi-infinite subsystems initially held at different temperatures and suddenly joined together \cite{BeDo16,VaMo16}, which corresponds to the so-called bipartition protocol. Note that before the introduction of generalised hydrodynamics, this setting could be studied only numerically \cite{GKSS05,SaMi13,AlHe14,BDVR16,ViIR17,PBPD17}, while analytical understanding was restricted to free systems \cite{ARRS99,AsPi03,AsBa06,PlKa07,LaMi10,EiRz13,DVBD13,Blocalquench, CoKa14,EiZi14,CoMa14,DeMV15,DLSB15,VSDH16,KoZi-1-17, Mint11,MiSo13} and conformal field theories \cite{SoCa08, CaHD08, BeDo12, DoHB14, BeDo15, BeDo16, ADSV16, BeDo16Review, DuSC17,SDCV17,EiBa17}. In the bipartition protocol, at large distances $x$ from the origin and times $t$, a different local quasi-stationary state emerges for each ``ray" $\zeta=x/t$. Accordingly, each physical quantity is associated with a non-trivial profile function of~$\zeta$. 

Systematic inspection of the profiles provides non-trivial information on the system and in particular on its quasi-particle content: in the XXZ Heisenberg chain, for instance, different bound states of spin excitations are associated to distinct points of non-analyticity in the profile functions of local observables \cite{PDCB17}. It is natural to wonder whether generalised hydrodynamics can be employed to study effects of separation in systems with non-trivial internal degrees of freedom. In general, however, signatures of this underlying structure are difficult to obtain. A simple example is provided by the recent study \cite{MBPC17}, where spreading of entanglement and mutual information after a global quench was analysed in a SU(3)-invariant spin chain by means of the techniques introduced in \cite{AlCa17}. Such spin chain is characterised by two species of quasi-particle excitations carrying different physical information and exhibits spin-charge separation at zero temperature. This structure, however, can not be observed in the spreading of information after the quench. While traces of different quasi-particles can be observed in the time-evolution, the resulting picture can not be distinguished by the one where there is a single species of quasi-particles forming bound states. In other words, all quasi-particles are observed to carry the same physical information. This scenario is also expected for the spreading of local correlations and is ultimately due to the general belief that separation effects vanish at finite energy density: they can not be observed beyond the regime of validity of the Luttinger theory~\cite{EFGK_book}.

In this work we investigate these questions by studying the transport properties of an interacting spinful Fermi gas in one dimension. Specifically, we consider the repulsive Yang-Gaudin model: a prominent example of nested-Bethe-ansatz-integrable system~\cite{korepin,takahashi} where spin and charge degrees of freedom are associated to distinct {species} of quasi-particles. Using the generalised hydrodynamics theory we show that for generic temperatures there is no trace of spin-charge separation in the profiles of local observables while a spin-charge-separated structure can be identified in the low-temperature regime. We develop an analytical low temperature expansion of the profiles of local observables, pinpointing qualitative features that could not be observed in models with a single quasi-particle species.

The paper is organised as follows. In Sec.~\ref{eq:model_and_quench} we introduce the Yang-Gaudin model of spinful Fermions and its thermodynamic Bethe ansatz description. In Sec.~\ref{sec:GHD} we describe the setting considered and review the generalised hydrodynamics treatment. In Sec.~\ref{sec:genT} we consider the profiles of local observables at generic temperatures and show that no sign of separation can be discerned. In Sec.~\ref{sec:lowT} we consider low temperatures and identify the signatures of spin-charge separation in the profiles of local observables. Sec.~\ref{sec:conclusions} contains our conclusions while a number of technical details of the low-temperature expansion are reported in the appendices.  

%%%%%%%%%%%%%%%%%%%
\section{The Yang-Gaudin model}%%
\label{eq:model_and_quench}%%%%
%%%%%%%%%%%%%%%%%%%

The Yang-Gaudin model describes a system of spin-1/2 fermions interacting via a repulsive delta-function potential. The Hamiltonian for the system in a finite volume $L$ reads as 
\begin{align}
\hat{H}=& -\int_{-{L/2}}^{{L/2}} {\rm d}x \left[\sum_{\alpha=\pm}\psi_\alpha^\dag(x) (\partial_x^2+A+\alpha h) \psi_\alpha(x)\right]\notag\\
&+ c\int_{-{L/2}}^{{L/2}} \!\!\!{\rm d}x \!\left[\sum_{\alpha,\beta=\pm} \psi^\dag_\alpha(x)\psi_\beta^\dag(x)\psi_\beta(x)\psi_\alpha(x) \right]\!\!.
\label{eq:Hamiltonian}
\end{align}
Here $h$ is an external magnetic field, $A$ is a chemical potential, and $c$ is the interaction strength. The fermionic fields $\psi^\dag_\alpha(x)$ and $\psi_\alpha(x)$ respectively create and destroy a fermion with spin $\alpha=\uparrow,\downarrow$ at position $x$ and fulfil the canonical anti-commutation relations 
\be
\{\psi^\dag_\alpha(x),\psi_\beta(y)\}=\delta_{\alpha,\beta}\delta(x-y)\,. 
\ee
In this paper we consider repulsive interactions, namely we take $c>0$. 

Considering the sector of $N$ fermions, $M$ of which have spin down, the Hamiltonian can be written in the following first-quantized form 
\begin{align}
  {\cal H}_{N,M}=&-\sum_{j=1}^{N} \left(\frac{\partial^{2}}{\partial x_j^{2}}+A\right) - h(N-2M)\notag\\
  &+ 2c \sum_{k \ne j=1}^N \delta(x_{k}-x_{j})\,.
  \label{eq:Hamiltonian1st}
\end{align}
The model is integrable and can be solved by nested Bethe ansatz~\cite{Yang67,Gaud67}. In this framework one writes the exact many-body wave-function $\psi_{\boldsymbol k, \boldsymbol \lambda}(x_{1},\dots,x_{N})$ of every eigenstate of ${\cal H}_{N,M}$. The eigenfunctions are labeled by a set of physical wave numbers $\boldsymbol k=\{k_{1},\dots,k_{N}\}$, corresponding to the physical particles, and a set of rapidities $\boldsymbol \lambda =\{\lambda_{1},\dots,\lambda_{M}\}$, corresponding to spin waves. Imposing periodic boundary conditions forces wave numbers and rapidities to fulfil the following algebraic equations  
\begin{align}
  e^{ik_{j}L}&=\prod_{\alpha=1}^{M}\frac{k_j-\lambda_{\alpha}+ic/2}{k_j-\lambda_{\alpha}-ic/2}, 
  \label{eq:physicalBetheEqProd} \\
  \prod_{j=1}^{N}\frac{\lambda_{\alpha}-k_{j}+ic/2}{\lambda_{\alpha}-k_{j}-ic/2}&=\prod_{\substack{\beta=1 \\ \beta\ne\alpha}}^{M}\frac{\lambda_\alpha-\lambda_{\beta}+ic}{\lambda_\alpha-\lambda_{\beta}-ic},
  \label{eq:auxiliaryBetheEqProd}
\end{align}
known as nested Bethe equations. These equations contain all the necessary information to study the thermodynamics %and the hydrodynamics 
of the system.

For large $L$, the solutions to \eqref{eq:physicalBetheEqProd}-\eqref{eq:auxiliaryBetheEqProd} follow the well-known string hypothesis~\cite{takahashi}, according to which the wave numbers $k_{j}$ are real, while the rapidities $\lambda_{\alpha}$ form patterns in the complex plane called strings. A $n$-string consists of $n$ rapidities distributed symmetrically around the real axis, with the $j$th rapidity in the string being 
\begin{equation}
  \lambda_{\alpha,j}^{n}=\lambda^{n}_{\alpha}+ i (n+1-2j)c'\,.
  \label{eq:lStringStructure}
\end{equation}
The real number $\lambda^{n}_{\alpha}$ is known as the string centre and satisfies a set of algebraic equations following from \eqref{eq:physicalBetheEqProd}-\eqref{eq:auxiliaryBetheEqProd}. 

In the thermodynamic limit 
\be
L \rightarrow \infty, \qquad N/L=\mathrm{const.}, \qquad M/L=\mathrm{const.},
\ee
the wave numbers $k_{j}$ and the string centres $\lambda^{n}_{\alpha}$ characterising a generic eigenstate become infinite in number and densely distributed. In this case, instead of using these quantities, it is useful to parametrise eigenstates using their ``root densities''
\begin{align}
  &\rho_1^{(1)}(k_{j})\sim \frac{1}{L} \frac{1}{k_{j+1}-k_{j}}\,,
  \label{eq:rRho1Definition}\\
  &\rho_n^{(2)}(\lambda^{n}_{\alpha})\sim \frac{1}{L} \frac{1}{\lambda^{n}_{\alpha+1}-\lambda^{n}_{\alpha}}.
  \label{eq:rRho2Definition}
\end{align}
In general, a large number of eigenstates correspond to the same set of densities. This fact is usually referred to by saying that the root densities specify a macrostate of the system, while the eigenstates of the Hamiltonian correspond to its microstates. Quantitatively, in a large finite volume $L$ there are $\sim\exp({L s_{\rm YY}[\boldsymbol \rho]})$ eigenstates approximately described by the same set of densities $\boldsymbol \rho =\{\rho_1^{(1)}, \rho^{(2)}_{n\phantom{1}}\}$. Here we introduced the Yang-Yang entropy density~\cite{takahashi} 
\be
  s_{\mathrm{YY}}[\boldsymbol \rho] = \sum_{r=1,2}\sum_{n=1}^{s_r}\,s^{(r)}_{n}[\boldsymbol \rho]\,, 
\ee
where $s_1=1$, $s_2=\infty$, and 
\be
    \begin{split}
    s^{(r)}_{n}[\boldsymbol \rho]&\equiv  \int_{-\infty}^{\infty}\! \mathrm d \lambda\, \rho^{(r)}_{n}\!(\lambda) \log\!\left[1\! +\! \rho^{(r)}_{n,\mathrm h}\!(\lambda)/\rho_{[n]}^{(r)}\!(\lambda)\right] \\
  & + \int_{-\infty}^{\infty}\! \mathrm d \lambda\, \rho^{(r)}_{n,\mathrm h}(\lambda) \log\!\left[1\! +\! \rho^{(r)}_{n}\!(\lambda)/\rho_{n,\mathrm{h}}^{(r)}\!(\lambda)\right].
  \end{split}
 \label{eq:sEntropyForOneSpeciesDefinition}
\ee
The hole densities $\rho^{(1)}_{1,\mathrm h}$, $\rho^{(2)}_{n,\mathrm h}$ appearing in these equations describe the densities of unoccupied momenta, and can be written in terms of $\rho^{(1)}_1$, $\rho^{(2)}_{n}$ as follows 
\begin{align}
  \rho^{(1)}_{1,\mathrm{h}}(\lambda)&=\frac{1}{2 \pi}-\rho^{(1)}_1(\lambda) + \sum_{m=1}^{\infty} a_{m} \star \rho^{(2)}_{m}(\lambda) ,
  \label{eq:repulsiveIntegralBGTPhysical}  \\
   \rho^{(2)}_{n,\mathrm{h}}(\lambda)&=a_n \star \rho^{(1)}_1(\lambda) -\!\!\!\! \sum_{m=1}^{\infty} (I_{nm}+A_{nm}) \star \rho^{(2)}_{m}(\lambda).
  \label{eq:repulsiveIntegralBGTAuxiliary}
\end{align}
Here, $f\star g$ denotes the convolution
\begin{equation}
 f \star g(\lambda) = \int_{-\infty}^{\infty} \mathrm d \mu f(\lambda - \mu) g(\mu),
  \label{eq:fConvolutionDefinition}
\end{equation}
while  
\begin{align}
  a_n(x) &\equiv  \frac{1}{\pi} \frac{2 n c}{(n c)^2+ 4 x^2},
   \label{eq:aIntegralConvolutionKernel}
\\
I_{nm}(x) &\equiv \delta_{nm}\delta(x)\,,
\\
  \begin{split}
  A_{nm}(x) &\equiv (1-\delta_{nm})a_{|n-m|}(x) + 2 a_{|n-m|+2}(x) \\
  &\quad + 2 a_{|n-m|+4}(x)+ \dots + 2 a_{n+m-2}(x)\\ &\quad + a_{n+m}(x).
\end{split}
  \label{eq:aIntegralConvolutionKernelBoundStates}
\end{align}
The equations \eqref{eq:repulsiveIntegralBGTPhysical}-\eqref{eq:repulsiveIntegralBGTAuxiliary} are known as Bethe-Gaudin-Takahashi equations and follow from \eqref{eq:physicalBetheEqProd}-\eqref{eq:auxiliaryBetheEqProd}.

Even if an exponential number of eigenstates correspond to the same densities $\rho^{(1)}_1$, $\rho^{(2)}_{n}$ it is generally accepted that the latter fully specify the thermodynamic limit of the expectation value of any physical observable in any of those eigenstates. For example, the expectation values of particle density, magnetisation density, and Hamiltonian density in an eigenstate characterised by $\boldsymbol \rho$ are given by 
\begin{align}
  n[\boldsymbol \rho] &=d^{(1)}[\boldsymbol \rho]=\int_{-\infty}^{\infty}\! \mathrm d \lambda\, \rho^{(1)}_1(\lambda)\,,\label{eq:nTBAstate}\\
  m[\boldsymbol \rho] &=d^{(1)}[\boldsymbol \rho]/2-d^{(2)}[\boldsymbol{\rho}]\,,\label{eq:mTBAstate}\\
  e[\boldsymbol \rho] &=\int_{-\infty}^{\infty}\! \mathrm d k\, \rho^{(1)}_1(k)\, e(k) - 2m[\boldsymbol \rho] h\,,
  \label{eq:eTBAstate}
\end{align}
where we introduced the bare energy ${e(\lambda)=\lambda^{2} - A}$ and
\be
d^{(2)}[\boldsymbol{\rho}]=\sum_{n=1}^{\infty}\, \int_{-\infty}^{\infty}\! \mathrm d \lambda\, \rho^{(2)}_{n}(\lambda)\,n\,.
\label{eq:second_species_density}
\ee 
Note that the particle density depends only on the density of particle 1 and $n[\boldsymbol \rho]/2-m[\boldsymbol \rho]$ depends only on the densities of particle 2.  Explicit formulae for the expectation values of the currents associated to these quantities can be obtained by generalising to the nested case the expressions found in Refs.~\cite{{CaDY16},{BCDF16}} and read as
\begin{align}
  j_n[\boldsymbol \rho] &=\int_{-\infty}^{\infty}\! \mathrm d \lambda\, v^{(1)}_{1}(\lambda) \rho^{(1)}_1(\lambda)\,,\label{eq:jnTBAstate}\\
   j_m[\boldsymbol \rho] & =j_n[\boldsymbol \rho]/2-\sum_{n=1}^{\infty}\, \int_{-\infty}^{\infty}\! \mathrm d \lambda\, v^{(2)}_{n}(\lambda) \rho^{(2)}_{n}(\lambda)\,n\,,\label{eq:jmTBAstate}\\
  j_e[\boldsymbol \rho] &=\int_{-\infty}^{\infty}\! \mathrm d k\, v^{(1)}_{1}(k)  \rho^{(1)}_1(k)\, (k^{2}-A) + 2 h j_m[\boldsymbol \rho]\,.
  \label{eq:jeTBAstate}
\end{align}
Here, the velocities of excitations $\boldsymbol v= \{v^{(1)}_1, v^{(2)}_{n\phantom{1}}\}$ on the state characterised by $\boldsymbol \rho$ are computed in terms of the root densities. In particular, they can be obtained by solving the system of linear integral equations \cite{BEL:prl}
\begin{align}
  v^{(1)}_{1}\rho^{(1)}_{1,\mathrm{t}}(\lambda)=&\frac{e'(\lambda)}{2 \pi} + \sum_{m=1}^{\infty} a_{m} \star v^{(2)}_{m}\rho^{(2)}_{m}(\lambda), 
\label{eq:vExcitationVelocitiesPhysical}  \\
  v^{(2)}_{n}\rho^{(2)}_{n,\mathrm{t}}(\lambda)=&a_n \star v^{(1)}_{1}\rho^{(1)}_{1}(\lambda)\notag\\
  &-\sum_{m=1}^{\infty} A_{nm} \star v^{(2)}_{m}\rho^{(2)}_{m}(\lambda),
  \label{eq:vExcitationVelocitiesAuxiliary}
\end{align}
where we introduced the total root densities
\be
\rho^{(r)}_{n,\mathrm{t}}(\lambda)\equiv \rho^{(r)}_{n \phantom{\mathrm{t}}}(\lambda)+\rho^{(r)}_{n, \mathrm{h}}(\lambda)\,.
\label{eq:rhotot}
\ee
Expressions similar to \eqref{eq:nTBAstate}-\eqref{eq:jeTBAstate} hold for the expectation values of density and current of each one of the local conserved charges related to the integrability of the model. For a generic charge $Q$ we have 
\begin{align}
  q[\boldsymbol \rho] &=\sum_{r=1,2}\sum_{n=1}^{s_r}\int_{-\infty}^{\infty}\! \mathrm d k\, \rho^{(r)}_{n}(k)\,q^{(r)}_{n}(k)\,,
  \label{eq:qTBAstate}\\
  j_q[\boldsymbol \rho] &=\sum_{r=1,2}\sum_{n=1}^{s_r}\int_{-\infty}^{\infty}\! \mathrm d k\, v^{(r)}_{n}(\lambda)  \rho^{(r)}_{n}(k)\,q^{(r)}_{n}(k)\,,
  \label{eq:jqTBAstate}
\end{align}
where $\{q^{(r)}_m (\lambda)\}$ are some known functions [which depend on $Q$].

\subsection{Homogeneous thermal state}
Let us consider the system in a homogeneous thermal state 
\be
\hat  \rho_{\rm th} = \frac{e^{-\beta { H}}}{{\rm tr}\left[e^{-\beta {H}}\right]}\,,
\label{eq:thermalstate}
\ee
where $H$ is the Hamiltonian \eqref{eq:Hamiltonian} and ${\beta^{-1}=T}$ is the temperature of the state. In this case, in the thermodynamic limit, one can adopt a microcanonical description and replace \eqref{eq:thermalstate} with %an appropriately chosen single eigenstate $\ket{{\rm mc}, T}$ of the Hamiltonian
a single eigenstate of the Hamiltonian in the expectation values of local observables. The root densities $\boldsymbol \rho_{T}$, corresponding to this particular eigenstate, are determined by minimising the free energy functional~\cite{takahashi}
\begin{equation}
  f_{T,A,h}[\boldsymbol \rho] = e[\boldsymbol \rho] - T\, s_{\mathrm YY}[\boldsymbol \rho]\,.
  \label{eq:fFreeEnergyDensity}
\end{equation}
This yields the following set of ``thermodynamic Bethe ansatz'' (TBA) equations  
\begin{align}
  \varepsilon^{(1)}_{1,T}(\lambda) = &e(\lambda) -h - T \sum_{n=1}^{\infty} a_{n} \star \log (1 + e^{-\varepsilon^{(2)}_{n,T}/T})(\lambda)\,, 
  \label{eq:repulsiveSaddlePointCoupled1} \\
 \varepsilon^{(2)}_{n, T}(\lambda) =& 2nh - T a_{n} \star \log (1 + e^{-\varepsilon^{(1)}_{1,T}/T})(\lambda) \notag\\
  &+ \sum_{m=1}^{\infty} A_{nm} \star \log (1 + e^{-\varepsilon^{(2)}_{m,T}/T})(\lambda)\,,
  \label{eq:repulsiveSaddlePointCoupled2}
\end{align}
where we introduced the thermal dressed energies ${\varepsilon^{(r)}_{m, T}(\lambda) \equiv T \log (\rho^{(r)}_{m, \mathrm{h},T}(\lambda)/\rho^{(r)}_{m,T}(\lambda))}$. We can write the velocities of the excitations on the thermal state in terms of these dressed energies 
\be
  v^{(r)}_{n,T}(\lambda)=\frac{\varepsilon^{(r)\prime}_{n,T}(\lambda)}{2\pi  \rho^{(r)}_{n,{\rm t},T}(\lambda)}\,,
\label{eq:veldef}
\ee
where the prime denotes the derivative with respect to $\lambda$ and $\rho^{(r)}_{n,{\rm t},T}(\lambda)$ are the total root densities (\emph{cf}. \eqref{eq:rhotot}) of the thermal state. Equation \eqref{eq:veldef} follows from the definition of the group velocity 
\be
v^{(r)}_{n,T}(\lambda)=\frac{\partial \varepsilon^{(r)}_{n,T}(\lambda)}{\partial p^{(r)}_{n,T}(\lambda)}\,,
\ee
where $p^{(r)}_{n,T}(\lambda)$ is the dressed momentum of the $n$-th bound state of the $r$-th species, and the observation 
\be
p^{(r)\prime}_{n,T}(\lambda)=2\pi\rho^{(r)}_{n,{\rm t},T}(\lambda)\,,
\label{eq:dressedmomentum}
\ee
see, \emph{e.g.}, Ref.~\cite{PDCB17}. Note that it is straightforward to see that \eqref{eq:veldef} is equivalent to the implicit definition given by the solution to \eqref{eq:vExcitationVelocitiesPhysical} and \eqref{eq:vExcitationVelocitiesAuxiliary}.

%%%%%%%%%%%%%%%%%%%%%%%%%%%%%%%%%%%%%%%%
\section{Bipartite setting and generalised hydrodynamic treatment}%%
\label{sec:GHD}%%%%%%%%%%%%%%%%%%%%%%%%%%%%%%%%
%%%%%%%%%%%%%%%%%%%%%%%%%%%%%%%%%%%%%%%%
The goal of this paper is to analyse the time-evolution of a bipartite state of the form  
\begin{equation}
\label{eq:ini_state}
\hat  \rho_0= \frac{1}{Z}e^{- \beta_L (\hat{H}_L -A_{L}\hat{N}_L-h_L\hat{M}_L)}\bigotimes e^{- \beta_R (\hat{H}_R -A_{R}\hat{N}_R-h_R\hat{M}_R)}\,,
\end{equation}
where $\hat{N}$, $\hat{M}$ are respectively the operators corresponding to the number of particles and the magnetization.
Here operators with the subscript $L$ and $R$ are defined by restricting the integrals of their density respectively to $x<0$ and $x>0$, while $Z$ is an appropriate normalisation constant. 

Using the same dephasing arguments~\cite{BS08} adopted in the case of homogeneous quantum quenches~\cite{QuenchReviews}, it is natural to conjecture that at large enough times the time-evolving state 
\be
\hat \rho(t)=e^{-i H t} \hat  \rho_0 e^{i H t}\,,
\ee
can be replaced by a quasi-stationary state in the expectation values of local observables~\cite{BeFa16}. Namely
\be
{\rm tr}\left[\hat \rho(t) \mathcal O(x)\right]\sim {\rm tr}\left[\hat \rho_{\rm S}(x,t) \mathcal O(0)\right],
\ee
where $x$ is the distance between the position of the local observable and the junction. Differently from the translationally invariant case, in the bipartite setting under exam the state $\hat \rho_{\rm S}(x,t)$ retains some ``weak'' dependence on $x$ and $t$. 

\begin{figure}[t]
\centering
\includegraphics[width=0.45\textwidth]{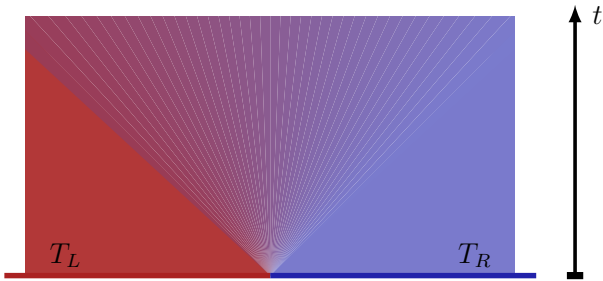}
\caption{Pictorial representation of the quench protocol. After the sudden junction of the two halves in two thermal states at temperatures $T_L$ and $T_R$ a non-trivial light-cone region emerges from the junction.}
\label{fig:sketch}
\end{figure}

In systems with a maximal velocity for the propagation of information $v_{\rm max}$ one can rigorously prove that observables at distance ${> t v_{\rm max}}$ (${< -t v_{\rm max}}$) from the junction are always described by the right (left) thermal state~\cite{BeFa16}. This means that the regions %on the left and on the right of the junction 
where the state locally looks thermal remain macroscopically large for every $t>0$ and play the role of effective thermal baths, while a third region emerges around the junction. This region is where non-trivial transport phenomena can be observed~\cite{BeDo16,VaMo16} (see Fig.~\ref{fig:sketch}). This picture does not strictly hold for non-relativistic field theories as the Yang-Gaudin model because there is no bound in the propagation of signals. In this case, however, an effective maximal velocity is set by the state~\eqref{eq:ini_state} as the fastest modes have an exponentially small occupation (a similar mechanism has been discussed in Refs.~\cite{Blocalquench,CaDY16}). This lightcone becomes sharp in the zero temperature limit where the role of lightcone velocity is played by the maximal Fermi velocity.  

This picture is generic and is believed to describe the dynamics from bipartite states in both integrable and non-integrable systems: the main difference between the two classes arises when one looks at how the non-trivial region around the junction scales with time. While for non-integrable models one expects a sub-ballistic scaling, for integrable systems the scaling is ballistic and the non-trivial region occupies a proper light cone emanating from the junction~\cite{BeFa16, CaDY16, BCDF16}. 

The ballistic scaling in integrable models is due to the presence of stable quasi-particle excitations that propagate information throughout the system with different velocities~\cite{BEL:prl}. At the leading order in $x$ and $t$, these quasi-particles can be assumed to move as free classical particles: the effects of the interactions are taken into account by letting their velocity depend on the state. The $x$- and $t$- dependence of expectation values of local observables can be explained by noting that varying $x$ and $t$ the observables ``measure" different amounts of quasi-particles from the two sides. This immediately explains the ballistic spreading of the light cone: the expectation value of a local observable moving away from the junction at fixed velocity $v$ starts to deviate from the thermal value as soon it starts to receive quasi-particles coming from the opposite side. This picture leads to a simple scaling form of the quasi-stationary state $\hat \rho_{\rm S}(x,t)= \hat \rho_{\rm S}(x/t)$. In other words everything depends only on the scaling variable $\zeta=x/t$, which is usually called ray.

At any fixed ray $\zeta$ the state $\hat \rho_{\rm S}(\zeta)$ can be characterised microcanonically in terms of a single set of root densities $\boldsymbol \rho_\zeta$, which can be thought of as the rapidity distributions of the quasi-particles in the state. The main result of Refs.~\cite{{CaDY16},{BCDF16}} was to determine the differential equation fulfilled by these root densities, which in our case reads as 
\be
\partial_t \rho^{(r)}_{n, \zeta}(\lambda)+\partial_x(v^{(r)}_{n, \zeta}(\lambda)\rho^{(r)}_{n,  \zeta}(\lambda)) = 0 \,.
\label{eq:continuity}
\ee 
Here $\boldsymbol v_\zeta= \{ v^{(1)}_{1, \zeta}, v^{(2)}_{n, \zeta}\}$ are the velocities of the quasi-particles excitations over the state $\boldsymbol \rho_\zeta$. Equation \eqref{eq:continuity} is in full agreement with the quasi-particle picture given above: the rapidity distributions change because the quasi-particles are moving with state-dependent velocities $\boldsymbol v_\zeta$.

Introducing the so-called filling functions 
\be
\vartheta^{(r)}_{n,\zeta}(\lambda)=\frac{\rho^{(r)}_{n,\zeta}(\lambda)}{\rho^{(r)}_{n,\mathrm{t},\zeta}(\lambda)}\,,
\label{eq:fillingfunction}
\ee
we can rewrite Eq.~\eqref{eq:continuity} as follows 
\begin{equation}
  \left( \zeta - v^{(r)}_{n,\zeta}(\lambda) \right) \partial_{\zeta} \vartheta^{(r)}_{n,\zeta}(\lambda) = 0\,.
  \label{eq:zHydrodynamicalEquation}
\end{equation}
This equation is implicitly solved by
\begin{equation}
  \begin{split}
    \vartheta^{(r)}_{n,\zeta}(\lambda) = \vartheta^{(r)}_{n, T_\mathrm L}(\lambda) \Theta_H (v^{(r)}_{n,\zeta}(\lambda) - \zeta)  & \quad \\ +\vartheta^{(r)}_{n,  T_\mathrm R}(\lambda) \Theta_H (\zeta-v^{(r)}_{n,\zeta}(\lambda))\,,
  \end{split}
  \label{eq:tHydrodynamicalSolutionTheta}
\end{equation}
where $\Theta_H (x)$ is the step function such that $\Theta_H (x)$ is non-zero and equal to one only if $x>0$, while $\vartheta^{(r)}_{n, T}$ are the filling functions of a thermal state at temperature $T$: they are written in terms of the thermal dressed energies $\varepsilon^{(r)}_{n, T}(\lambda)$ (\emph{cf}.~\eqref{eq:repulsiveSaddlePointCoupled1}-\eqref{eq:repulsiveSaddlePointCoupled2}) as follows 
\be
\vartheta^{(r)}_{n, T}(\lambda)=\frac{1}{1+e^{\varepsilon^{(r)}_{n, T}(\lambda)/T}}\,.
\ee
The solution~\eqref{eq:tHydrodynamicalSolutionTheta} is implicit because the velocities $\boldsymbol v_{\zeta}$ depend on $\boldsymbol \vartheta_\zeta=\{\vartheta^{(1)}_{1,\zeta},\vartheta^{(2)}_{m,\zeta}\}$, and to determine the filling functions one normally needs to resort on a numerical solution by iteration: one starts with an initial guess for the velocities  $\boldsymbol v_\zeta$, finds $\boldsymbol \vartheta_{\zeta}$ using \eqref{eq:tHydrodynamicalSolutionTheta} and then uses Eqs.~\eqref{eq:repulsiveIntegralBGTPhysical}-\eqref{eq:repulsiveIntegralBGTAuxiliary} and Eqs.~\eqref{eq:vExcitationVelocitiesPhysical}-\eqref{eq:vExcitationVelocitiesAuxiliary} to find the new velocities. As we will see in the following, an exception is represented by the low-temperature case, where an explicit analytic solution can be achieved.  

Once the filling functions $\boldsymbol \vartheta_\zeta$ are known one can determine the root densities %$\{\rho^{(r)}_{n,\zeta}\}$ 
by solving \eqref{eq:repulsiveIntegralBGTPhysical}-\eqref{eq:repulsiveIntegralBGTAuxiliary} and use them to compute the profiles of conserved charges densities and currents. In particular, in this paper we will always consider the profiles of the densities \eqref{eq:nTBAstate}-\eqref{eq:eTBAstate} and the currents \eqref{eq:jnTBAstate}-\eqref{eq:jeTBAstate}.   

\begin{figure*}
	\centering
	\includegraphics[scale=0.8]{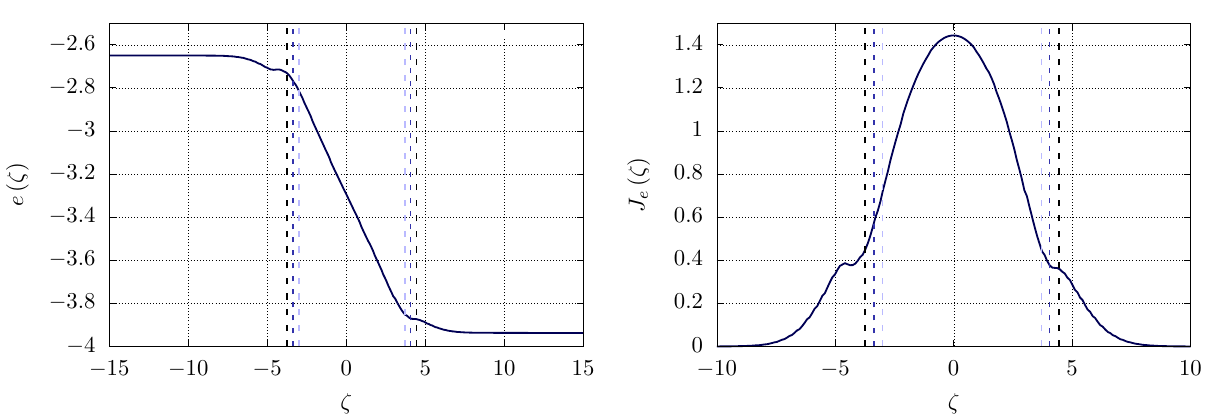}
	\caption{Profiles of energy density (Left) and current (Right) as a function of the ray $\zeta=x/t$ at infinite times after a quench from the bipartite state~\eqref{eq:ini_state}. These plots are obtained by a numerical solution of Eq.~\eqref{eq:tHydrodynamicalSolutionTheta}.
		The parameters of the Hamiltonian \eqref{eq:Hamiltonian1st} are chosen as $c=1$, $h=0.1$, $A=5$, while the parameters of the initial state \eqref{eq:ini_state} are $\beta_{\mathrm{L}}=0.5$, $A_L=-2$, $h_L=0.1$ and $\beta_{\mathrm{R}}=5$, $A_R=0$, $h_R=0.3$.}
	\label{fig:HT1}
\end{figure*}

%%%%%%%%%%%%%%%%%%%%%%%
\section{Profiles at generic temperatures}%%
\label{sec:genT}%%%%%%%%%%%%%%%
%%%%%%%%%%%%%%%%%%%%%%%

\begin{figure*}
  \centering
  \includegraphics[scale=0.8]{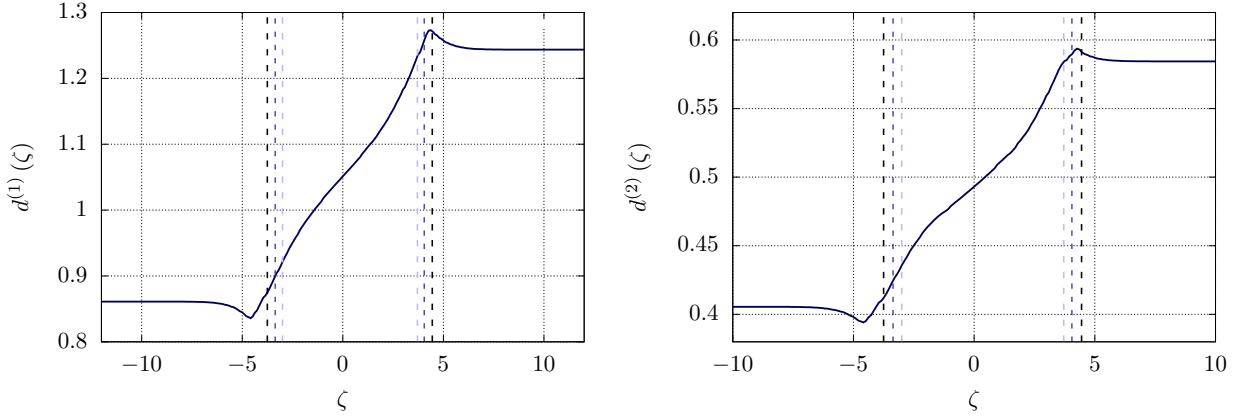}
  \caption{Profiles of the density of particles of the first (Left) and second species (Right), as a function of the ray $\zeta=x/t$ at infinite times after a quench from the bipartite state~\eqref{eq:ini_state}. The parameters of the Hamiltonian \eqref{eq:Hamiltonian1st} are chosen as $c=1$, $h=0.1$, $A=5$, while the parameters of the initial state \eqref{eq:ini_state} are $\beta_{\mathrm{L}}=0.5$, $A_L=-2$, $h_L=0.1$ and $\beta_{\mathrm{R}}=5$, $A_R=0$, $h_R=0.3$.}
  \label{fig:HT2}
\end{figure*}

Let us start by studying the profiles of the densities and currents at generic temperatures by numerical evaluation of the implicit solution~\eqref{eq:tHydrodynamicalSolutionTheta}. Our results are reported in Figs.~\ref{fig:HT1} and \ref{fig:HT2}, where two representative examples are considered. 

As it was observed in \cite{PDCB17}, non-analyticities in the profiles are expected at the rays $\zeta^{\pm}_n$ corresponding to the maximal (minimal) velocity of the $n$-spin bound states, which are identified as the solution to the equations
\bea
\zeta_n^{-}&=&{\rm min}\left[v_{n,\zeta_n^{-}}(\lambda)\right]\,,\\
\zeta_n^{+}&=&{\rm max}\left[v_{n,\zeta_n^{+}}(\lambda)\right]\,.
\eea
These rays are reported in Figs.~\ref{fig:HT1} and \ref{fig:HT2} as vertical lines. We see that these mathematical non-analyticities are difficult to observe from our numerical solution. We verified that this is due to the fact that for the thermal quenches we have considered the occupation numbers of the second species are small in correspondence of the modes with the maximal (minimal) velocity.

Note, instead, that the quasi-particles associated to the physical rapidities do not have a maximal velocity: as we already mentioned, it is effectively set by the occupation numbers of the initial state, which are exponentially vanishing for large quasi-momenta. From Figs.~\ref{fig:HT1} and \ref{fig:HT2} one can identify the effective maximum (minimum) velocity of the first species of quasi-particles in correspondence of the point where the profiles start to deviate from a constant value.

In general the profiles are determined by both species of quasi-particles. This is true for any observable, including, for instance the density of the second species displayed in Fig~\ref{fig:HT2}. Indeed, even if the expectation value of the latter only involves the spin rapidity distribution function, \emph{cf}. \eqref{eq:second_species_density}, the Bethe equations \eqref{eq:repulsiveIntegralBGTPhysical} and \eqref{eq:repulsiveIntegralBGTAuxiliary} couple the two species, so that variations of the charge root density brings about a variation of the spin one. More physically this means that whenever there is a finite density of particles of the two species, the way in which the velocity of one of the species is dressed depends also on the density of the other species. As a consequence of this coupling the profiles of all charges and currents display the same qualitative structure of those observed in the XXZ spin-1/2 chain~\cite{PDCB17}, and in fact cannot be distinguished from those of a model with a single species of quasi-particles forming bound states. We conclude that no sign of the nested structure of the system emerges in this setting.

%%%%%%%%%%%%%%%%%%%%
\section{Profiles at low temperatures}%
\label{sec:lowT}%%%%%%%%%%%%
%%%%%%%%%%%%%%%%%%%%

Let us now move our attention to the case where the temperatures $T_{\mathrm{L}}$ and $T_{\mathrm{R}}$ are very small and show that in this case some signatures of spin-charge separation can be observed. We develop a low temperature expansion of the profiles by generalising to the multiple species case the treatment put forward in Ref.~\cite{BePi17} for the XXZ spin-1/2 chain. 

To warm up, let us consider the low temperature expansion of the (constant) profiles in the homogeneous case, namely when $T_{\mathrm{R}} =  T_{\mathrm{L}}\equiv T$. This helps developing some concepts and strategies that are directly applied to the inhomogeneous case. The calculations in the low-temperature regimes, will always be performed for initial states \eqref{eq:ini_state} with $A_L=h_L=A_R=h_R=0$ (while the parameters $A$ and $h$ of the driving Hamiltonian \eqref{eq:Hamiltonian} are left arbitrary).

\subsection{Low-temperature expansion in the homogeneous system}
\label{sub:homogeneousExpansion}

\begin{figure}
	\centering
	\includegraphics[scale=0.65]{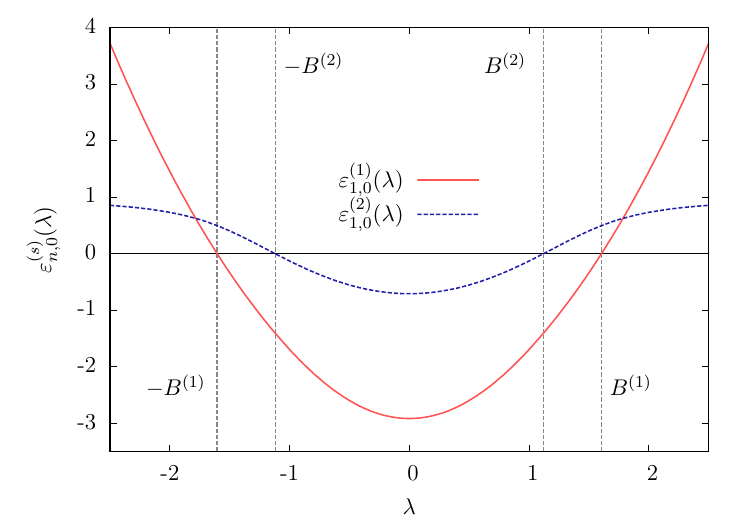}
	\caption{Ground-state pseudoenergies \eqref{eq:ezeroTemperatureTBA1}--\eqref{eq:ezeroTemperatureTBA2} at ${c=1}$, ${h=0.5}$, and ${A=2}$. The values $B^{(1)}$ and $B^{(2)}$, corresponding to the Fermi rapidities, are highlighted with vertical dotted lines. The pseudoenergies are monotonous in $\lambda^{2}$.}
	\label{fig:zeroTemperatureEpsilon}
\end{figure}

We start by summarising Takahashi's procedure to obtain the leading order in the low temperature description of the model \cite{Taka71}, obtaining the ground state expectation values in \eqref{eq:nTBAstate}-\eqref{eq:jeTBAstate}. The ground state root densities and velocities are found by taking the zero temperature limit of \eqref{eq:repulsiveIntegralBGTPhysical}-\eqref{eq:repulsiveIntegralBGTAuxiliary}, \eqref{eq:repulsiveSaddlePointCoupled1}-\eqref{eq:repulsiveSaddlePointCoupled2}, and \eqref{eq:vExcitationVelocitiesPhysical}-\eqref{eq:vExcitationVelocitiesAuxiliary}. The limit is taken by starting from the equations \eqref{eq:repulsiveSaddlePointCoupled1}-\eqref{eq:repulsiveSaddlePointCoupled2}, which determine the thermal dressed energies. Noting that the dressed energies of the bound states $\varepsilon^{(2)}_{n \ge 2,0}(\lambda)$ are always positive (see Appendix \ref{sub:noBoundStates}), we can write the zero temperature limit as follows~\cite{Taka71} 
\begin{align}
  \varepsilon^{(1)}_{1,0}(\lambda) &= \lambda^{2}-h-A + a_{1}\ast \varepsilon^{(2)}_{1,0}(\lambda)\bigr |_2,
  \label{eq:ezeroTemperatureTBA1} \\
  \begin{split}
  \varepsilon^{(2)}_{1,0}(\lambda) &= 2 h + a_{1}\ast \varepsilon^{(1)}_{1,0}(\lambda)\bigr |_1 - a_{2}\ast \varepsilon^{(2)}_{1,0}(\lambda)\bigr |_2\,,
\end{split}
  \label{eq:ezeroTemperatureTBA2} 
\end{align}
where we introduced the notation
\be
f\ast g(\lambda)\bigr |_r\equiv \int_{-B^{(r)}}^{B^{(r)}}\! \mathrm{d} \mu\, f(\lambda-\mu) g(\mu)\,.
\label{eq:GSconvolution}
\ee
In writing \eqref{eq:ezeroTemperatureTBA1}-\eqref{eq:ezeroTemperatureTBA2} we used that $\varepsilon^{(r)}_{1,0}(\lambda)$ are monotonic functions of $\lambda^2$ (see Fig. \ref{fig:zeroTemperatureEpsilon} for an illustration and the Appendix of~\cite{Taka71} for a proof). The intervals $[-B^{(r)},B^{(r)}]$ are the intervals in which $\varepsilon^{(r)}_{1,0}(\lambda)$ are negative. In other words, the rapidities $B^{(r)}$ are the ``Fermi rapidities'' bounding the Fermi sea at zero temperature.

\begin{figure}
	\centering
	\includegraphics[width=3.5in]{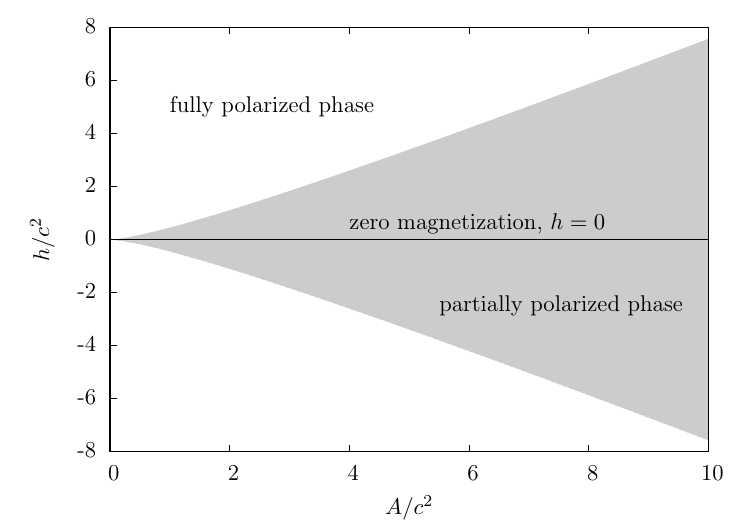}
	\caption{Ground-state phase diagram of the repulsive Yang--Gaudin model. The critical line between the partially and fully polarised phase is given by \eqref{eq:criticalHEquationExplicitIntegral}.}
	\label{fig:fig1}
\end{figure}

These equations can be used to determine the ground state phase diagram of the system, which we report in Fig. \ref{fig:fig1}. The main features of the diagram are as follows. At zero magnetic field $B^{(2)}=\infty$ and the magnetisation is zero. Increasing the magnetic field $B^{(2)}$ becomes finite and the magnetisation increases. There is a critical magnetic field $h_{\mathrm{crit}}(A)$ above which $B^{(2)}=0$ and the ground state becomes fully polarised. Since $\varepsilon^{(2)}_{1,0}(\lambda)$ has its global minimum in zero, this critical field is found by imposing $\varepsilon^{(2)}_{1,0}(0)=0$, which yields
\begin{equation}
  \begin{split}
  0 &= 2 h_{\mathrm{crit}} + \frac{1}{ 2\pi} \bigg[ 2 c \sqrt{A+h_{\mathrm{crit}}}\! - \!\left( 4A + c^{2}\! +\! 4 h_{\mathrm{crit}} \right) \, \\ & \qquad \times \arctan\! \left( \frac{ 2 \sqrt{A + h_{\mathrm{crit}}}}{ c} \right) \bigg].
  \label{eq:criticalHEquationExplicitIntegral}
\end{split}
\end{equation}
The critical line $h_{\mathrm{crit}}(A)$ implicitly defined here is shown in the phase diagram of Fig. \ref{fig:fig1}.

Let us now take the zero temperature limit of \eqref{eq:repulsiveIntegralBGTPhysical}-\eqref{eq:repulsiveIntegralBGTAuxiliary}, and \eqref{eq:vExcitationVelocitiesPhysical}-\eqref{eq:vExcitationVelocitiesAuxiliary}, which determine the root densities and the velocities of excitations. The zero temperature limit of \eqref{eq:repulsiveIntegralBGTPhysical}-\eqref{eq:repulsiveIntegralBGTAuxiliary} reads as 
\begin{align}
  \rho^{(1)}_{1,{\rm t},0}(\lambda) &= \frac{1}{2 \pi} +  a_{1}\ast \rho^{(2)}_{1,{\rm t},0}(\lambda)\bigr |_2\,, 
  \label{eq:rZeroTemperatureDensities1} \\ 
  \begin{split}
    \rho^{(2)}_{1,{\rm t},0}(\lambda) &= a_{1}\star \rho^{(1)}_{1,{\rm t},0}(\lambda)\bigr |_1 - a_{2}\ast \rho^{(2)}_{1,{\rm t},0}(\lambda)\bigr |_2\,.
  \label{eq:rZeroTemperatureDensities2}
  \end{split}
\end{align}
To obtain the system \eqref{eq:rZeroTemperatureDensities1}-\eqref{eq:rZeroTemperatureDensities2} we used
\be
\rho^{(r)}_{n, T}(\lambda)=\frac{\rho^{(r)}_{n,\mathrm{t}, T}(\lambda)}{1+e^{\varepsilon^{(r)}_{n, T}(\lambda)/T}}.
\ee
This implies that in the zero temperature limit only the root densities for $n=1$ are not vanishing. In particular they are non-zero only in the intervals $[-B^{(r)},B^{(r)}]$ where $\rho^{(r)}_{1, 0}(\lambda)=\rho^{(r)}_{1,\mathrm{t}, 0}(\lambda)$. Proceeding analogously we find the following expression for the zero temperature limit of \eqref{eq:vExcitationVelocitiesPhysical}-\eqref{eq:vExcitationVelocitiesAuxiliary} 
\begin{align}
 &\!\rho^{(1)}_{1,{\rm t},0} v^{(1)}_{1,0}(\lambda) = \frac{\lambda}{\pi} +  a_{1} \ast \rho^{(2)}_{1,{\rm t},0} v^{(2)}_{1,0}(\lambda)\bigr |_1, 
  \label{eq:rZeroTemperatureVelocities1} \\
    &\!\rho^{(2)}_{1,{\rm t},0} v^{(2)}_{1,0}(\lambda) = a_{1}\ast \rho^{(1)}_{1,{\rm t},0}v^{(1)}_{1,0}(\lambda)\bigr |_1\notag\\
    &\qquad\qquad\qquad - a_{2}\ast \rho^{(2)}_{1,{\rm t},0}v^{(2)}_{1,0}(\lambda)\bigr |_2.
  \label{eq:rZeroTemperatureVelocities2}
\end{align}
Note that the above equations imply $\rho^{(r)}_{1,\mathrm{t},0}(-\lambda) =\rho^{(r)}_{1,\mathrm{t},0}(\lambda) $ while $v^{(r)}_{1,0}(-\lambda)=-v^{(r)}_{1,0}(\lambda)$.

The ground state expectation values are found plugging the solutions of \eqref{eq:rZeroTemperatureDensities1}-\eqref{eq:rZeroTemperatureVelocities2} into \eqref{eq:nTBAstate}-\eqref{eq:jeTBAstate}. In particular, one immediately finds that the expectation values of the currents \eqref{eq:jnTBAstate}-\eqref{eq:jeTBAstate} are zero because the integrand is odd in $\lambda$. This can be immediately understood using reflection symmetry: even charges have odd currents and their thermal expectation value is always zero because the Hamiltonian is even. Since we are taking the limit $T\rightarrow0$ after the thermodynamic limit this property continues to hold in the limit. 

Let us now determine the first finite-$T$ correction to those values. We start by introducing the short-hand notation
\be
\delta f(\lambda) = f(\lambda)-f_0(\lambda)\,, 
\ee
to denote the difference between a quantity and its ground state value. At small but finite temperature, the modes with ${\varepsilon^{(n)}_{m, 0}(\lambda)>0}$ are still exponentially suppressed. Therefore we can describe the state using the thermal dressed energies $ \varepsilon^{(r)}_{1,T}(\lambda)$: the bound states of the particles of the second species can be neglected from the integral equations. Following Refs.~\cite{BePi17,Taka73}, we find the leading finite-$T$ corrections by writing a linear system of integral equations for $\delta \varepsilon^{(n)}_{1, T}(\lambda)$. The derivation is reported in Appendix~\ref{sec:homogeneousLowTemperatureExpansionAppendix} while the result reads as 
\begin{align}
  \delta \varepsilon^{(1)}_{1, T}(\lambda) &= \frac{ \pi^2 T^2 }{6\, \varepsilon^{(2)\prime}_{1,0}(B^{(2)})}U^{(1)}(\lambda) + O(T^{4}), 
  \label{eq:FirstPseudoEnergyCorrection1} \\
  \delta \varepsilon^{(2)}_{1, T}(\lambda) &= \frac{ \pi^2 T^2 }{6\, \varepsilon^{(2)\prime}_{1,0}(B^{(2)})}U^{(2)}(\lambda) + O(T^{4}),
  \label{eq:FirstPseudoEnergyCorrection2}
\end{align}
where the functions $U^{(r)}(\lambda)$ satisfy
\begin{align}
    &\!\!\!U^{(1)}(\lambda) = d_U^{(1)}(\lambda) + a_{1}\ast U^{(2)}(\lambda)\bigr |_2,
   \label{eq:repU1Definition}\\
    &\!\!\!U^{(2)}(\lambda) =  d_U^{(2)}(\lambda)+ a_{1}\ast U^{(1)}(\lambda)\bigr |_1- a_{2}\ast U^{(2)}(\lambda)\bigr |_2\,.
    \label{eq:repU2Definition}
\end{align}
The driving functions appearing here can be written in the following convenient form 
\begin{align}
 &d_U^{(1)}(\lambda)=a_1\ast z^{(2)}_U(\lambda)\bigr |_2,\label{eq:driving1}\\ 
 &d_U^{(2)}(\lambda)=a_1\ast z^{(1)}_U(\lambda)\bigr |_1-a_2\ast z^{(2)}_U(\lambda)\bigr |_2,\label{eq:driving2}
\end{align}
where we introduced 
\begin{align}
 z^{(r)}_U(\lambda)=&\sum_{\sigma=\pm} z^{(r)}_{U,\sigma}(\lambda)\,,\label{eq:driving3}
\end{align}
and finally 
\begin{align}
z^{(r)}_{U,\sigma}(\lambda)&\equiv z^{(r)}(C^{(r)}_{U,\sigma},D^{(r)}_{U,\sigma}, \lambda) \label{eq:driving4}\\
&=\, C^{(r)}_{U,\sigma}\delta(\lambda-\sigma B^{(r)-})+D^{(r)}_{U,\sigma}\delta'(\lambda-\sigma B^{(r)-})\,.\notag
\end{align}
Here $\delta(x)$ and $\delta'(x)$ are respectively the Dirac delta and its first derivative, ${B^{(r)-}=B^{(r)}e^{-\epsilon}}$ for infinitesimal ${\epsilon>0}$, and the constants are given by ${C^{(2)}_{U,\sigma}=-1}$,  $C^{(1)}_{U,\sigma}={\varepsilon^{(2)\prime}_{1,0}(B^{(2)})}/{\varepsilon^{(1)\prime}_{1,0}(B^{(1)})}$, and ${D^{(r)}_{U,\sigma}=0}$. 

Let us now sketch how to find the correction to the total root densities, while we refer to Appendix~\ref{sec:homogeneousLowTemperatureExpansionAppendix} for the full derivation. Neglecting the contribution of bound states, we can rewrite \eqref{eq:repulsiveIntegralBGTPhysical}-\eqref{eq:repulsiveIntegralBGTAuxiliary} on a thermal state as follows 
\begin{align}
  &\rho^{(1)}_{1,\mathrm{t},T}(\lambda)=\frac{1}{2 \pi} + a_{1} \star \vartheta^{(2)}_{1,T}\rho^{(2)}_{1,\mathrm{t},T}(\lambda), \label{eq:firstrholowT} \\
  &\rho^{(2)}_{1,\mathrm{t},T}(\lambda)=a_1 \star \vartheta^{(1)}_{1,T}\rho^{(1)}_{1,\mathrm{t},T}(\lambda) - a_2\star \vartheta^{(2)}_{1,T}\rho^{(2)}_{1,\mathrm{t},T}(\lambda).\label{eq:secondrholowT}
\end{align}
These equations feature many Sommerfeld-like integrals of the form 
\be
\int_{-\infty}^{\infty}\! \mathrm{d} \lambda\,\,\, \vartheta^{(r)}_{1,T}(\lambda) f(\lambda)
\label{eq:integral}
\ee   
with appropriate functions $f(\lambda)$. The expansion of such integrals is carried out in Appendix A of Ref.~\cite{BePi17} (there the expansion is carried out in the gapless phase of the XXZ spin-1/2 chain but it applies to all TBA-solvable models where $\varepsilon^{(r)}_{1,0}(\lambda)$ has two symmetric zeros) and reads as
\begin{align}
 I^{(r)}_{f}=\int_{-B^{(r)}}^{B^{(r)}}\!\!\!\!\! \mathrm{d} \lambda\,\left[1+ \frac{\pi^2T^2}{6} z_I^{(r)}(\lambda)\right]f(\lambda)+O(T^4)\,,
\label{eq:exp}
\end{align} 
where $z_I^{(r)}(\lambda)$ is of the form \eqref{eq:driving3}-\eqref{eq:driving4} with the coefficients
\begin{align}
    D^{(r)}_{I,\sigma} &=   \frac{-\sigma}{ (\varepsilon^{(r)\prime}_{1,0}(B^{(r)}))^2}, \label{eq:exp1}\\
   \frac{C^{(2)}_{I,\sigma}}{D^{(2)}_{I,+} } &= \!\!\left[ \frac{ \varepsilon^{(2)\prime\prime}_{1,0}(B^{(2)})}{\varepsilon^{(2)\prime}_{1,0}(B^{(2)})} +   U^{(2)}(B^{(2)})\right]\!\!,
  \label{eq:exp2} \\
    \frac{C^{(1)}_{I,\sigma}}{D^{(1)}_{I,+} }&=   \!\!\left[\frac{ \varepsilon^{(1)\prime\prime}_{1,0}(B^{(1)})}{\varepsilon^{(1)\prime}_{1,0}(B^{(1)})} +  \tilde U^{(1)}(B^{(1)})\right]\!\!,
  \label{eq:exp3}
\end{align}
where we introduced 
\be
\tilde U^{(1)}(B^{(1)})=U^{(1)}(B^{(1)})\frac{ \varepsilon^{(1)\prime}_{0}(B^{(1)})}{\varepsilon^{(2)\prime}_{0}(B^{(2)})}\,. 
\ee
Using this expression we can turn \eqref{eq:firstrholowT}-\eqref{eq:secondrholowT} into a set of linear integral equations for the corrections to the total root densities. At the leading order in $T$ we find 
\begin{align}
  \delta\rho^{(1)}_{1,{\mathrm t},T}(\lambda) &= \frac{\pi^{2}T^{2}}{6 } R^{(1)}(\lambda) + O(T^{4}),
  \label{eq:lowTemperatureDensitiesRDefinitions1} \\
  \delta \rho^{(2)}_{1, {\mathrm t}, T}(\lambda)  &= \frac{\pi^{2}T^{2}}{6 } R^{(2)}(\lambda) + O(T^{4}),
  \label{eq:lowTemperatureDensitiesRDefinitions2}
\end{align}
where $R^{(r)}(\lambda)$ fulfil the system \eqref{eq:repU1Definition}-\eqref{eq:repU2Definition} with drivings $d_R^{(r)}(\lambda)$. The drivings are given by \eqref{eq:driving1}-\eqref{eq:driving2}, where $z_U^{(r)}(\lambda)$ is replaced by $z_{R,I}^{(r)}(\lambda)$ of the form \eqref{eq:driving3}-\eqref{eq:driving4} and coefficients
\begin{align}
    D^{(r)}_{R,\sigma} &=   -\sigma \frac{\rho^{(r)}_{1, \mathrm t,0}(B^{(r)})}{ (\varepsilon^{(r)\prime}_{1,0}(B^{(r)}))^2},  \label{eq:lowTemperatureDensities1}\\
   \frac{C^{(2)}_{R,\sigma}}{D^{(2)}_{R,-} } &= \!\!\left[ \frac{\rho^{(2)\prime}_{1,0}(B^{(2)})}{\rho^{(2)}_{1,\mathrm t,0}(B^{(2)}) } -\frac{ \varepsilon^{(2)\prime\prime}_{1,0}(B^{(2)})}{\varepsilon^{(2)\prime}_{1,0}(B^{(2)})} -   U^{(2)}(B^{(2)})\right]\!\!,
  \label{eq:lowTemperatureDensities2} \\
    \frac{C^{(1)}_{R,\sigma}}{D^{(1)}_{R,-} }&=   \!\!\left[ \frac{\rho^{(1)\prime}_{1,0}(B^{(1)})}{\rho^{(1)}_{1,\mathrm t,0}(B^{(1)})} -  \frac{ \varepsilon^{(1)\prime\prime}_{1,0}(B^{(1)})}{\varepsilon^{(1)\prime}_{1,0}(B^{(1)})} -  \tilde U^{(1)}(B^{(1)})\right]\!\!.
  \label{eq:lowTemperatureDensities3}
\end{align}
Similarly, we obtain the following expression for the corrections to the velocities 
\begin{align}
  \delta\rho^{(1)}_{1, {\mathrm t}, T} v^{(1)}_{1, T}(\lambda) &= \frac{\pi T^{2}}{6 } W^{(1)}(\lambda) + O(T^{4}),
  \label{eq:lowTemperatureVelocitiesWDefinitions1} \\
  \delta \rho^{(2)}_{1, {\mathrm t}, T} v^{(2)}_{1, T}(\lambda)  &= \frac{\pi T^{2}}{6 } W^{(2)}(\lambda) + O(T^{4}).
  \label{eq:lowTemperatureVelocitiesWDefinitions2}
\end{align}
The functions $W^{(r)}(\lambda)$ fulfil the system \eqref{eq:repU1Definition}-\eqref{eq:repU2Definition} with drivings $d_W^{(r)}(\lambda)$ again given by \eqref{eq:driving1}-\eqref{eq:driving2}. This time the function $z_U^{(r)}(\lambda)$ is replaced by $z_{W,I}^{(r)}(\lambda)$ of the form \eqref{eq:driving3}-\eqref{eq:driving4} and coefficients
\begin{align}
  D^{(r)}_{W,\sigma} &=    \frac{-1}{ \varepsilon^{(r)\prime}_{1,0}(B^{(r)})}, \label{eq:lowTemperatureWCD1}\\
  \frac{C^{(2)}_{W,\sigma}}{D^{(2)}_{W,-} } &=  \sigma U^{(2)}(B^{(2)}), \label{eq:lowTemperatureWCD2}\\ 
  \frac{C^{(1)}_{W,\sigma}}{D^{(1)}_{W,-} }&=  \sigma \tilde U^{(1)}(B^{(2)}). \label{eq:lowTemperatureWCD3}
\end{align}
We see that the driving functions $d_W^{(r)}(\lambda)$ are odd, implying that also the functions $W^{(r)}(\lambda)$ are odd. 

Using these expressions we can finally write the first non-trivial corrections to the densities  \eqref{eq:nTBAstate}-\eqref{eq:eTBAstate} at finite temperature. In particular, let us consider the first finite-temperature correction to the energy density
\begin{align}
\delta e &=\int_{-\infty}^{\infty}\! \mathrm d \lambda\, \left[\delta \rho^{(1)}_{1,T}(\lambda)\, e(\lambda) + 2h \sum_{n=1}^{\infty} \delta\rho^{(2)}_{n,T}(\lambda) n\right]\notag\\
&=  \int_{-\infty}^{\infty}\! \mathrm d \lambda\, \left[\delta \rho^{(1)}_{1,T}(\lambda)\, e(\lambda) + 2h \delta \rho^{(2)}_{1,T}(\lambda)\right],
\end{align}
where in the second step we neglected exponentially small corrections in $1/T$. Using the expansion \eqref{eq:exp} and the expressions \eqref{eq:lowTemperatureDensitiesRDefinitions1}-\eqref{eq:lowTemperatureDensitiesRDefinitions2} we have 
\begin{align}
  \delta e &= \frac{\pi^{2}T^{2}}{6 } \int_{-\infty}^{\infty}\! \mathrm d \lambda\, (R^{(1)}(\lambda)+z_{R,I}^{(1)}(\lambda))  \cdot e(\lambda)\notag\\
  &\quad+\frac{\pi^{2}T^{2}}{6 } \int_{-\infty}^{\infty}\! \mathrm d \lambda\, (R^{(2)}(\lambda)+z_{R,I}^{(2)}(\lambda))\cdot 2h\notag\\
&= \frac{\pi^{2}T^{2}}{6 }  \!\!\sum_{r=1,2}\sum_{\sigma=\pm} \left\{C^{(r)}_{R,\sigma}\varepsilon^{(r)}_{1,0}(\sigma B^{(r)})\!-\!D^{(r)}_{R,\sigma}\varepsilon^{(r)\prime}_{1,0}(\sigma B^{(r)})\right\}\notag\\
&= \frac{\pi^2 T^{2}}{3}  \!\!\sum_{r=1,2} \!\!\frac{\rho^{(r)}_{1, \mathrm t,0}(B^{(r)})}{ \varepsilon^{(r)\prime}_{1,0}(B^{(r)})}= \frac{\pi T^{2}}{6}   \!\!\sum_{r=1,2} \frac{1}{v^{(r)}_{1, 0}(B^{(r)})}.\label{eq:edenscorr}
\end{align}
%Here we introduced $w_R^{(r)}(\lambda)$, of the form \eqref{eq:driving3}-\eqref{eq:driving4} with coefficients $C^{(r)\prime}_{R,\sigma}=C^{(r)}_{R,\sigma}$ and $D^{(r)\prime}_{R,\sigma}=-D^{(r)}_{R,\sigma}$ (\emph{cf}. Eqs.~\eqref{eq:lowTemperatureDensities1}-\eqref{eq:lowTemperatureDensities3}). 
Here in in the second step we used the identity 
\begin{align}
  & \sum_{r=1,2} \int_{-\infty}^{\infty}\! \mathrm d \lambda\, (R^{(r)}(\lambda)+z_{R,I}^{(r)}(\lambda))q^{(r)}(\lambda)\notag\\
&= \!\!\sum_{r=1,2}\sum_{\sigma=\pm} \left\{C^{(r)}_{R,\sigma}f^{(r)}_q(\sigma B^{(r)})\!-\!D^{(r)}_{R,\sigma}f^{(r)\prime}_q(\sigma B^{(r)})\right\}
\label{eq:identity}
\end{align} 
where the functions $f^{(r)}_{q}(\lambda)$ fulfil 
\begin{align}
 &f^{(1)}_{q}(\lambda)= q^{(1)}(\lambda) +a_{1}\ast f^{(2)}_{q}(\lambda)\bigr |_2,
 \label{eq:fFunctionDefinition1} \\
  &f^{(2)}_{q}(\lambda)= q^{(2)}(\lambda) + a_{1}\ast f^{(1)}_{q}(\lambda)\bigr |_1-a_{2}\ast f^{(2)}_{q}(\lambda)\bigr |_2.
 \label{eq:fFunctionDefinition2} 
\end{align}
The identity \eqref{eq:identity} is proven by inverting the integral system for $R^{(r)}(\lambda)$, see Appendix~\ref{sec:homogeneousLowTemperatureExpansionAppendix} for the detailed proof. In the third step of \eqref{eq:edenscorr} we noted that, for $q^{(1)}(\lambda)=e(\lambda)$ and $q^{(2)}(\lambda)=h$, equations \eqref{eq:fFunctionDefinition1}-\eqref{eq:fFunctionDefinition2} coincide with \eqref{eq:ezeroTemperatureTBA1}-\eqref{eq:ezeroTemperatureTBA2}, namely $f^{(r)}_{q}(\lambda)=\varepsilon^{(r)}_{1,0}(\lambda)$. In the last step we used the definition \eqref{eq:veldef} of the velocities.  

The finite temperature correction \eqref{eq:edenscorr} agrees with that of two independent conformal field theories (CFTs)~\cite{A:CFT, BCN:CFT} with central charge equal to one and velocity of light respectively equal to $v^{(1)}_{1, 0}(B^{(1)})$ and $v^{(2)}_{1, 0}(B^{(2)})$: the ``Fermi velocities" of the two components. This is in accordance with the well-known fact that the low energy description of \eqref{eq:Hamiltonian} is in terms of two decoupled CFTs. 

Before concluding our analysis of the homogeneous case we note that one can repeat this calculation for the other densities \eqref{eq:nTBAstate} and \eqref{eq:mTBAstate} finding an analogous result: finite temperature corrections proportional to $T^2$ and written as a sum of the spin and charge component. We also note that the corrections to the currents are all zero, because $W^{(r)}(\lambda)$ and $d_W^{(r)}(\lambda)$ are both odd. This is again in accordance with the fact that the expectation values of the currents vanish in a thermal state.

\begin{figure*}
	\centering
	\includegraphics[scale=0.8]{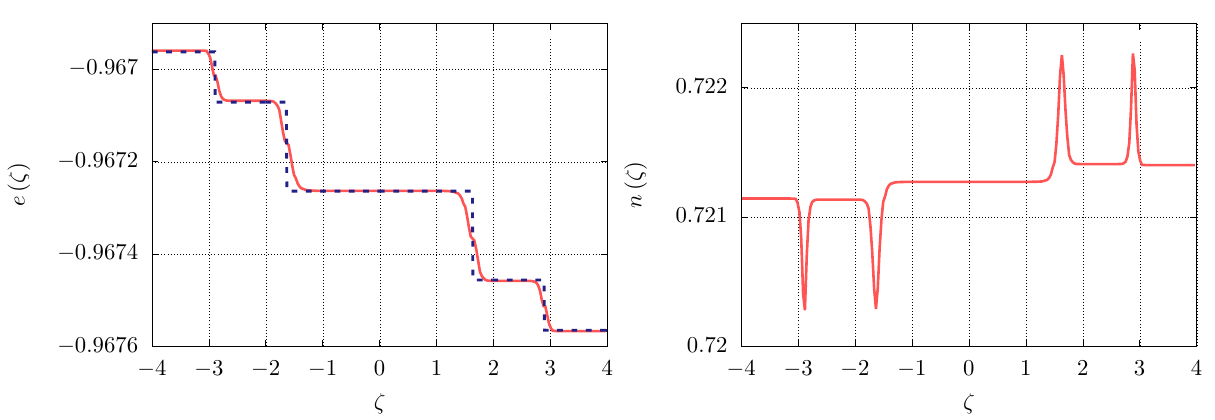}
	\caption{Profiles of energy (Left) and particle (Right) densities as a function of the ray $\zeta=x/t$ at infinite times after a quench from the bipartite state~\eqref{eq:ini_state}. Full lines are obtained as a numerical solution of Eq.~\eqref{eq:tHydrodynamicalSolutionTheta} while dashed lines are the result of the analytical low-temperature expansion~\eqref{eq:GeneralEnergyDensity}-\eqref{eq:GeneralEnergyCurrent}. The parameters of the Hamiltonian are set to $c=1$, $h=0.5$ and $A=2$, while those of the initial state \eqref{eq:ini_state} are $\beta_{\mathrm{L}}=25$ and $\beta_{\mathrm{R}}=50$ (with $A_R=h_R=0$, $A_L=h_L=0$).}
	\label{fig:lowTemperaturePlateaus_I}
\end{figure*}

\subsection{Low-temperature expansion in the inhomogeneous system}
\label{sub:inhomogeneousExpansion}

Let us now move to consider the inhomogeneous case. More precisely, we take $T_{\mathrm{L}} \equiv T$, $T_{\mathrm{R}} \equiv \frak r T$ and expand for small $T$ (with $A_R=h_R=0$, $A_L=h_L=0$). At the lowest order we again find that all the relevant quantities attain their ground state value.  To find the first finite temperature corrections we use the form of the implicit solution \eqref{eq:tHydrodynamicalSolutionTheta}. Both terms in \eqref{eq:tHydrodynamicalSolutionTheta} are multiplied by a filling function $\vartheta^{(r)}_{1,T}$ of a homogeneous thermal state. We can then use the reasoning above and conclude that all bound states of the spin rapidities can be neglected, as they give exponentially small corrections in $T$. We are then left to consider reduced systems of integral equations as \eqref{eq:firstrholowT}-\eqref{eq:secondrholowT} for $\rho^{(r)}_{1, {\mathrm t}, \zeta}$ (the same system, with a different driving, is found for $v^{(r)}_{1,\zeta}\rho^{(r)}_{1,\mathrm{t},\zeta}$). In order to find the leading low-temperature contribution to this system we need to expand integrals of the form  
\begin{equation}
  I^{(r)}_{\zeta,f}= \int_{-\infty}^{\infty}\! \mathrm{d} \lambda\, \vartheta^{(r)}_{1,\zeta}(\lambda) f^{(r)}(\lambda),
  \label{eq:iIntegralToCompute}
\end{equation}
for some appropriate functions $f^{(r)}(\lambda)$. The expansion of \eqref{eq:iIntegralToCompute} up to $O(T^3)$ is thoroughly carried out in Appendix~A of Ref.~\cite{BePi17}; below we discuss the main features. As a function of $\zeta$, two different regions appear where the integral behaves very differently. For rays $O(T^0)$ far from the lightcones of both particle species, i.e., when
\begin{equation}
  \lim_{T\rightarrow 0}\,|\zeta-v^{(r)}_{1,0}(B^{(r)})| \ne 0,
  \label{eq:lFarFromLightconeCondition}
\end{equation}
the expansion reads as 
\be
I^{(r)}_{\zeta,f} = \int_{-B^{(r)}}^{B^{(r)}}\! \mathrm d \lambda \, f^{(r)}(\lambda)\left[1+\frac{ \pi^{2} T^{2}}{6} z_{I,\zeta}^{(r)}(\lambda)\right]\,,
\label{eq:LowTemperatureIntegralSeries1}
\ee
where we neglected $O(T^4)$ and introduced 
\bea
z_{I,\zeta}^{(r)}(\lambda)&=& z_{I}^{(r)}(\lambda) \Theta_H [-v^{(r)}_{1,0}(B^{(r)})-\zeta] \nonumber\\ 
   &+&( \frak r^2 z_{I,-}^{(r)}(\lambda)+ z_{I,+}^{(r)}(\lambda))\,  \Theta_H [v^{(r)}_{1,0}(B^{(r)})-|\zeta|] \nonumber\\
   &+& \frak r^2 z_{I}^{(r)}(\lambda)\Theta_H [\zeta-v^{(r)}_{1,0}(B^{(r)})]\,.
   \label{eq:zFunction}
\eea
 The functions $z^{(r)}_{I}$ and $z^{(r)}_{I,\pm}$ are respectively of the form \eqref{eq:driving3} and \eqref{eq:driving4} and are specified by the coefficients \eqref{eq:exp1}--\eqref{eq:exp3}.
In words: far away from the light cones, the $O(T^{2})$ correction to the integral is piecewise constant, consisting of three plateaux. In the region 
\be
\zeta\pm v^{(r)}_{1,0}(B^{(r)})\sim O(T)\,,
\label{eq:region2}
\ee
we instead have 
\be
I^{(r)}_{\zeta,f}=\!\int_{-B^{(r)}}^{B^{(r)}}\!\!\!\!{\rm d}\lambda\,\, f^{(r)}(\lambda) \left[1+\frac{\pi^2 T (1-\frak r^2)}{6}z_{II}^{(r)}(\lambda,\zeta)\right]\,,
\label{eq:LowTemperatureIntegralSeries2}
\ee
where we neglected $O(T^2)$ and introduced the function 
\be
z_{II}^{(r)}(\lambda,\zeta)=\!\!\sum_{\sigma=\pm}\! z_{II,\sigma}^{(r)}(\lambda) {\cal D}_{\frak r} \!\!\left[\frac{\zeta -\sigma v^{(r)}_{1,0}(B^{(r)})}{ v^{(r)}_{1,0}(B^{(r)})) T |m^{(r)}_*|^{-1} } \right].
\ee
In this equation we introduced the effective mass 
\be
m^{(r)}_*=\frac{\partial^2 \varepsilon^{(r)\,\prime}_{1,0}(\lambda)}{\partial p^{(r)}_{1,T}(\lambda)^2}\Big |_{\lambda=B^{(r)} }=\frac{\varepsilon^{(r)\,\prime}_{1,0}(B^{(r)})v^{(r)}_{1,0}(B^{(r)})}{v^{(r)\,\prime}_{1,0}(B^{(r)})}\,,
\ee
where $ p^{(r)}_{n,T}(\lambda)$ is the dressed momentum \eqref{eq:dressedmomentum}, and the function 
\be
{\cal D}_{\frak r} [z]\equiv\frac{6\log(1+e^{z})}{\pi^2(1-{\frak r}^2)} -\frac{6{\frak r}\log(1+e^{z/{\frak r}})}{\pi^2(1-{\frak r}^2)}\,.
\label{eq:frakDr}
\ee
This function is positive and peaked around $z=0$, in particular we have 
\be
\lim_{T\rightarrow0^+}\frac{1}{T}{\cal D}_{\frak r} [z/T]=\delta(z)\,.
\ee
Finally, we also introduced the functions $z_{II,\sigma}^{(r)}(\lambda) $, which are of the form \eqref{eq:driving4} with coefficients ${C^{(r)}_{II,\sigma}=\sigma\, {\text{sgn}(v^{0\,\prime}_1(B^{(r)}))}/{\varepsilon_1'(B^{(r)})}}$ and $D^{(r)}_{II,\sigma}=0$.  

We see that in the region \eqref{eq:region2} the first correction to the integral \eqref{eq:iIntegralToCompute} is $O(T)$. As a function of $\zeta$ it has a peaked form described by the function ${\cal D}_{\frak r} [z]$. 

Using the expansions \eqref{eq:LowTemperatureIntegralSeries1} and  \eqref{eq:LowTemperatureIntegralSeries2} it is possible to write linear equations for the first order corrections to $\rho^{(r)}_{1, {\mathrm t}, \zeta} $ and $\rho^{(r)}_{1, {\mathrm t}, \zeta} v^{(r)}_{1, \zeta}(\lambda) $ as we did in the previous section for the homogeneous case. Proceeding as in \eqref{eq:edenscorr}, these expressions can be used to find the first correction to the profiles of local observables. Since the procedure is very similar to the one outlined in the previous section, we omit it here (it is reported for completeness in Appendix~\ref{sec:inhomogeneousLowTemperatureExpansionAppendix}). In the region \eqref{eq:lFarFromLightconeCondition} the result up to $O(T^{2})$ reads as
\begin{widetext}
\begin{align}
  \delta q(\frak r,\zeta)&= \frac{ \pi^2 T^{2} }{6} \left\{\sum_{r=1}^{2}\nu^{(r)}(\frak r,\zeta) \sum_{\sigma=\pm} \left[ C_{R,\sigma}^{(r)} f^{(r)}_{q}(\sigma B^{(r)})-D^{(r)}_{R,\sigma} f^{(r)\prime}_{q}(\sigma B^{(r)})) \right]\right\}, 
  \label{eq:EvenChargeStationaryValues}
  \\
  \delta j_{q}(\frak r,\zeta)&= \frac{ \pi^2 T^{2}}{6} \left\{ \sum_{r=1}^{2}\omega^{(r)}(\frak r,\zeta)  \sum_{\sigma=\pm}  \left[C_{W,\sigma}^{(r)} f^{(r)}_{q}(\sigma B^{(r)})-D^{(r)}_{W,\sigma} f^{(r)\prime}_{q}(\sigma B^{(r)})) \right] \right\},
  \label{eq:EvenCurrentStationaryValues}
\end{align}
where $\nu^{(r)}(\frak r,\zeta)$ and $\omega^{(r)}(\frak r,\zeta)$ are piecewise constant functions
\begin{align}
  \nu^{(r)}(\frak r,\zeta) &\equiv 1 + \frac{\frak r^{2}-1}{2} \bigg[\Theta_H\!\!\left[\zeta + v^{(r)}_{1,0}(B^{(r)})\right] +\Theta_H\!\!\left[\zeta-v^{(r)}_{1,0}(B^{(r)})\right]\bigg],
  \label{eq:nuFunctionDefinition} \\
  \omega^{(r)}(\frak r,\zeta) &\equiv \frac{1-\frak r^{2}}{2}\,\Theta_H\!\!\left[v^{(r)}_{1,0}(B^{(r)})-|\zeta|\right],
  \label{eq:omegaFunctionDefinition}
\end{align}
and $f^{(r)}_{q}(\lambda)$ is defined via the integral equations \eqref{eq:fFunctionDefinition1}-\eqref{eq:fFunctionDefinition2}. The coefficients $C^{(r)}_{R,\sigma},D^{(r)}_{R,\sigma},C^{(r)}_{W,\sigma},D^{(r)}_{W,\sigma}$ are defined in \eqref{eq:lowTemperatureDensities1}--\eqref{eq:lowTemperatureDensities3} and \eqref{eq:lowTemperatureWCD1}--\eqref{eq:lowTemperatureWCD3}. Note that the charge density expression \eqref{eq:EvenChargeStationaryValues} is similar to the homogeneous case \eqref{eq:edenscorr}--\eqref{eq:identity}, the only difference being the appearance of the profile function $\nu^{(r)}(\frak r,\zeta)$. 

In the region \eqref{eq:region2}, instead, the profiles up to $O(T)$ read as 
\begin{align}
 q(\frak r,\zeta)&=\sum_{r=1}^{2}\sum_{\sigma=\pm}\sigma\frac{\pi T \text{sgn}(v^{{(r)}\,\prime}_{1,0}(B^{(r)}))}{12  v^{(r)}_{1,0}(B^{(r)})}(1-\frak r^2)f_q( B^{(r)}){\cal D}_{\frak r} \!\!\left[\frac{\zeta -\sigma v^{(r)}_{1,0}(B^{(r)})}{ v^{(r)}_{1,0}(B^{(r)})) T |m^{(r)}_*|^{-1} } \right], 
  \label{eq:EvenChargeStationaryValues2}
  \\
  j_{q}(\frak r,\zeta)&=\sum_{r=1}^{2}\sum_{\sigma=\pm}\frac{\pi T \text{sgn}(v^{{(r)}\,\prime}_{1,0}(B^{(r)}))}{12}(1-\frak r^2)f_q( B^{(r)}){\cal D}_{\frak r} \!\!\left[\frac{\zeta -\sigma v^{(r)}_{1,0}(B^{(r)})}{ v^{(r)}_{1,0}(B^{(r)})) T |m^{(r)}_*|^{-1} } \right]\,,
  \label{eq:EvenCurrentStationaryValues2}
\end{align}
\end{widetext}
where ${\cal D}_{\frak r}[z]$ is defined in \eqref{eq:frakDr}. These results hold for the profile of any reflection symmetric charge density and the relative current, namely for conserved charges $Q$ characterised by bare charges $q^{(r)}_{n}(\lambda)$ (\emph{cf}.~\eqref{eq:qTBAstate}-\eqref{eq:jqTBAstate}) which are even functions of $\lambda$. 
\begin{figure*}
	\centering
	\includegraphics[scale=0.8]{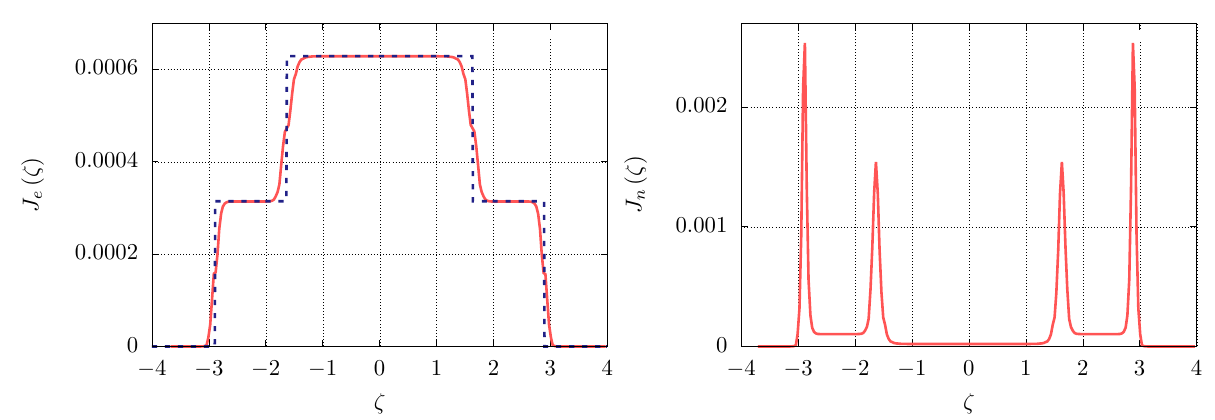}
	\caption{Profiles of energy (Left) and particle (Right) currents as a function of the ray $\zeta=x/t$ at infinite times after a quench from the bipartite state~\eqref{eq:ini_state}. Full lines are obtained as a numerical solution of Eq.~\eqref{eq:tHydrodynamicalSolutionTheta} while dashed lines are the result of the analytical low-temperature expansion~\eqref{eq:GeneralEnergyDensity}-\eqref{eq:GeneralEnergyCurrent}. The parameters of the Hamiltonian are set to $c=1$, $h=0.5$ and $A=2$, while those of the initial state \eqref{eq:ini_state} are $\beta_{\mathrm{L}}=25$ and $\beta_{\mathrm{R}}=50$ (with $A_R=h_R=0$, $A_L=h_L=0$).}
	\label{fig:lowTemperaturePlateaus_II}
\end{figure*}

The form of the profiles \eqref{eq:EvenChargeStationaryValues}-\eqref{eq:EvenCurrentStationaryValues} and  \eqref{eq:EvenChargeStationaryValues2}-\eqref{eq:EvenCurrentStationaryValues2} can be reproduced by considering two decoupled non-linear Luttinger liquids in the spirit of Ref.~\cite{BePC18}. In particular, in the region \eqref{eq:lFarFromLightconeCondition} the structure is that given by two decoupled linear Luttinger liquids, namely two decoupled CFTs. The profiles are piecewise constant functions of $\zeta$, and change value every time that the absolute value of the ray equals one of the two Fermi velocities $v^{(r)}_{1,0}(B^{(r)})$. This generically results in a five step form, as it can be seen from Figs.~\ref{fig:lowTemperaturePlateaus_I} and \ref{fig:lowTemperaturePlateaus_II}. In the following, we discuss our results displayed in these figures.

Let us first focus on the important profiles of the energy density and current. In this case we have $f^{(r)}_{q}(\lambda)=\varepsilon^{(r)}_{1,0}(\lambda)$, so that $f^{(r)}_{q}(B^{(r)})=\varepsilon^{(r)}_{1,0}(B^{(r)})=0$. Plugging this in \eqref{eq:EvenChargeStationaryValues}-\eqref{eq:EvenCurrentStationaryValues} and  \eqref{eq:EvenChargeStationaryValues2}-\eqref{eq:EvenCurrentStationaryValues2} we see that the leading contribution to the profiles comes only from the region~\eqref{eq:lFarFromLightconeCondition} and reads as 
\begin{align}
  {e}(\zeta,\frak r)&= {e_0}+ \frac{\pi T^{2}}{6}\,\sum_{r=1}^{2}\frac{ \nu^{(r)}(\frak r,\zeta)}{v^{(r)}_{1,0}(B^{(r)})}, 
  \label{eq:GeneralEnergyDensity}\\ 
  j_{e}(\zeta,\frak r)&= \frac{ \pi T^{2}}{6}\, \sum_{r=1}^{2} \omega^{(r)}(\frak r,\zeta).
  \label{eq:GeneralEnergyCurrent}
\end{align}
This result is compared with the numerical solution of the generalised hydrodynamic equations in the left panels of Figs.~\ref{fig:lowTemperaturePlateaus_I} and \ref{fig:lowTemperaturePlateaus_II}. It is of interest to write down explicitly these functions for $\zeta=0$, which corresponds to the celebrated non-equilibrium steady state (NESS)
\begin{align}
  {e}(0,\frak r)- {e_0}&= \frac{\pi  (T^{2}_{\rm L}+T^2_{\rm R})}{12}\,\sum_{r=1}^{2}\frac{ 1}{v^{(r)}_{1,0}(B^{(r)})}, 
  \label{eq:NessEnergyDensity}\\ 
  j_{e}(0,\frak r)&= \frac{ \pi (T^{2}_{\rm L}-T^2_{\rm R})}{6}\,.
  \label{eq:NessEnergyCurrent}
\end{align}
We note that the dependence on the temperatures of the two halves is the one expected from the CFT analysis of the bipartition protocol~\cite{BeDo12,BeDo16}: the results correspond to the sum of two CFTs with central charge equal to $1$. This is of course consistent with the description in terms of two decoupled Luttinger liquids. 

Next, we compare these profiles with those of other generic observables, focusing in particular on the quasi-particle density and current. These are displayed in the right panels of Figs.~\ref{fig:lowTemperaturePlateaus_I} and \ref{fig:lowTemperaturePlateaus_II}, and immediately show visible qualitative differences. Indeed, in the transition region between two plateaux, the non-linearity of the dispersion relations (quantified by $1/m^{(r)}$) becomes relevant~\cite{BePC18}: the profiles have a peaked form determined by the function ${\cal D}_{\frak r} [z]$. Note that the width of the peak around $\zeta=v^{(r)}_{1,0}(B^{(r)})$ depends on the species $r$. This form of the profiles was first described in \cite{BePC18} for a single Luttinger liquid, where it was shown to be a remarkable example of \emph{universality} beyond the linear Luttinger approximation \cite{ImGl09-science}. The calculations presented in this work give a non-trivial test of the validity of this prediction in the case of interacting nested systems.

The peculiar structure of the profiles described above, five-steps with peaks in correspondence of the transitions, 
gives an indication of a spin-charge separation. Indeed, observing profiles with this structure one can argue that they are produced by two decoupled non-linear Luttinger liquids. Note, however, that the local observables couple the two theories: it is impossible to find an observable sensitive to a single Luttinger liquid only. This means that one cannot determine the physical content of the two separated degrees of freedom and distinguish them by looking at the profiles: the signature of spin-charge separation that one can obtain from the profiles is only an indirect one. Still these are qualitatively different from the profiles that one would obtain in the low-temperature regime of non-nested systems, such as the XXZ Heisenberg chain \cite{BePi17,BePC18}, and highlight that the model has two species of quasi-particles.

%%%%%%%%%%%%%
\section{Conclusions}%%
\label{sec:conclusions}%%
%%%%%%%%%%%%%

We have considered the transport properties of a one-dimensional spinful Fermi gas, after junction of two semi-infinite sub-systems held at different temperatures. We have analysed quantitatively the space-time profiles of local observables emerging at large distances $x$ from the junction and times $t$, as a function of $\zeta=x/t$. By employing the generalised hydrodynamic approach, we have shown how an indirect signature of spin-charge separation emerges in the transport properties of the model at low temperatures. In this regime, profiles are qualitatively different from those that can be obtained in non-nested systems, and a description in terms of two decoupled Luttinger liquids can be employed, yielding exact results in the limit $T\to 0$. Among other results, we have seen how the universal predictions of \cite{BePC18} are recovered analytically from the generalised hydrodynamic equations. The calculations presented in this work generalise to nested systems those of \cite{BePi17}, and together  provide a comprehensive analysis of low-temperature transport properties in non-relativistic integrable systems.

\section*{Acknowledgments}

The authors thank Fabian Essler for useful discussions. B.B. acknowledges support from the ERC under the Advanced grant number 694544 (OMNES). P.C. acknowledges support from the ERC under the Consolidator grant  number 771536 (NEMO). Part of this work has been carried out during the workshop ``Quantum Paths''  at  the  Erwin  Schr\"odinger  International Institute (ESI) in Vienna.

%%%%%%%%%%%%%%%%%%%%%%%%%%%%%%%%%%
%% Appendices
%\clearpage
\onecolumngrid
\appendix
\section{Low-temperature expansion of the homogeneous TBA}
\label{sec:homogeneousLowTemperatureExpansionAppendix}

In this appendix we derive of the formulae of the homogeneous low temperature expansion appearing in Section~\ref{sub:homogeneousExpansion}. 

\subsection{The absence of bound states}
\label{sub:noBoundStates}

In the low temperature limit the bound states of the second species are absent. These states have a strictly positive pseudoenergy $\varepsilon^{(2)}_{n}(\lambda)$, and therefore their occupation numbers
\begin{equation}
  \vartheta^{(2)}_{n,T}(\lambda) = \frac{1}{1 + e^{-\varepsilon^{(2)}_{n,T}/T}} \quad (n \ge 2),
  \label{eq:occupationNumbersOfBoundStates}
\end{equation}
are exponentially suppressed. 

The positivity of the pseudoenergies $\varepsilon^{(2)}_{n,T}(\lambda)$ is seen from the decoupled version \cite{takahashi} (Chapter 12) of the TBA equations \eqref{eq:repulsiveSaddlePointCoupled1}--\eqref{eq:repulsiveSaddlePointCoupled2}
\begin{align}
  \varepsilon^{(1)}_{1,T} &= (\lambda^{2}\!-\!A) - T \big[ r \star \log(1+e^{-\varepsilon^{(1)}_{1,T}/T})  %\\& \quad 
  - s \star \log(1+e^{\varepsilon^{(2)}_{1,T}/T})\big],
  \label{eq:repulsiveSaddlePoint1} \\
  \varepsilon^{(2)}_{1,T} &= T \bigg[ s \star \log \left( \frac{1+e^{\varepsilon^{(2)}_{2,T}/T}}{1+e^{-\varepsilon^{(1)}_{1,T}/T}} \right) \bigg],
  \label{eq:repulsiveSaddlePoint2} \\
  \varepsilon^{(2)}_{n\ge 2,T} &= T \! \bigg[ s \star \log\! \left( (1\!+\! e^{\varepsilon^{(2)}_{n-1, T}/T})(1\! +\! e^{\varepsilon^{(2)}_{n+1,T}/T}) \right) \bigg]\!,
  \label{eq:repulsiveSaddlePoint3} \\
  \lim_{n \rightarrow \infty} \frac{\log \varepsilon^{(2)}_{n,T}}{n} &= \frac{2h}{T}\,,
  \label{eq:repulsiveSaddlePoint4}
\end{align}
where 
\be
s(x)\equiv \frac{1}{2c}{\rm sech}\left(\frac{\pi x}{c}\right)\,,\quad\qquad\qquad\qquad r(x)= a_1\star s(x)\,.
\ee
Equation \eqref{eq:repulsiveSaddlePoint3} shows that the pseudoenergies of the bound states are always positive. Indeed, both $s(x)$ and ${\log(1+ e^{\varepsilon^{(2)}_{n,T}(x)/T})}$ are strictly positive functions. 

\subsection{Expansion of the pseudoenergy}
\label{sub:homogeneousPseudoEnergyExpansion}
In the following we present the derivation of the expansions \eqref{eq:FirstPseudoEnergyCorrection1}--\eqref{eq:FirstPseudoEnergyCorrection2} of the pseudoenergies. We start from the TBA equations \eqref{eq:repulsiveSaddlePointCoupled1}--\eqref{eq:repulsiveSaddlePointCoupled2}
\begin{align}
  \begin{split}
    \varepsilon^{(1)}_{1,T}(\lambda) = &e(\lambda) - h  -T \sum_{m=1}^{\infty} a_{m} \star \log (1 \! + \! e^{-\varepsilon^{(2)}_{m,T}/T})(\lambda),
  \end{split} 
  \label{eq:repulsiveSaddlePointCoupled1Appendix} 
  \\
  \varepsilon^{(2)}_{n, T}(\lambda) =& 2nh - T a_{n} \star \log (1 + e^{-\varepsilon^{(1)}_{1,T}/T})(\lambda) \notag
  + T\, \sum_{m=1}^{\infty} A_{nm} \star \log (1 + e^{-\varepsilon^{(2)}_{m,T}/T})(\lambda).
  \label{eq:repulsiveSaddlePointCoupled2Appendix}
\end{align}
Since the pseudoenergy of the bound states is always positive (see \eqref{eq:repulsiveSaddlePoint3}), the $n \ge 2$ terms of the sums are exponentially suppressed at small temperatures. After subtracting the ground state equations \eqref{eq:ezeroTemperatureTBA1}--\eqref{eq:ezeroTemperatureTBA2}, we have
\begin{align}\\
  \begin{split}
    \delta \varepsilon^{(1)}_{1,T}(\lambda) &= - T\,  a_{1} \star \log (1   + e^{-\varepsilon^{(2)}_{1,T}/T})(\lambda) - \int_{-B^{(2)}}^{B^{(2)}}\! \mathrm{d} \mu\, a_{1}(\lambda-\mu) \varepsilon^{(2)}_{1,0}(\mu) + O(e^{-A_1/T}),
 \end{split} 
  \label{eq:ezeroTemperatureTBA1Appendix} \\
  \begin{split}
    \delta \varepsilon^{(2)}_{1,T}(\lambda) &= - T\,  a_{1} \star \log (1   + e^{-\varepsilon^{(1)}_{1,T}/T})(\lambda) - \int_{-B^{(1)}}^{B^{(1)}}\! \mathrm{d} \mu\, a_{1}(\lambda-\mu) \varepsilon^{(1)}_{1,0}(\mu) \\
    & \quad + T\,  a_{2} \star \log (1   + e^{-\varepsilon^{(2)}_{1,T}/T})(\lambda) + \int_{-B^{(2)}}^{B^{(2)}}\! \mathrm{d} \mu\, a_{2}(\lambda-\mu) \varepsilon^{(2)}_{1,0}(\mu) 
    + O(e^{-A_2/T}),
  \end{split}
  \label{eq:ezeroTemperatureTBA2Appendix}
\end{align}
with some $\alpha^{(r)} > 0$.
Now we plug in the relation
\begin{equation}
  \log (1 + e ^{-f(\lambda)}) = - f^-(\lambda) + \log (1 + e ^{-|f(\lambda)|}),
  \label{eq:lLog1PlusExpTrick}
\end{equation}
where 
\begin{equation}
f^{\pm}(\lambda) = (f(\lambda)\pm|f(\lambda)|)/2,
\end{equation}
obtaining
\begin{align}
\begin{split}
  \delta \varepsilon^{(1)}_{1,T}(\lambda)\! &=\!- \int_{-B^{(2)}}^{B^{(2)}} d \mu\, a_{1}(\lambda-\mu) \delta \varepsilon^{(2)}_{1,T}(\mu)-\int_{-B^{(2)\prime}}^{-B^{(2)}} d \mu\, a_{1}(\lambda-\mu) \varepsilon^{(2)}_{1,T}(\mu)- \int_{B^{(2)}}^{B^{(2)\prime}} d \mu\, a_{1}(\lambda-\mu) \varepsilon^{(2)}_{1,T}(\mu) 
  \\ &+ T \int_{-\infty}^{\infty}\! d\mu\, a_{1}(\lambda-\mu) \log (1\!+\! e^{-|\varepsilon^{(2)}_{1,T}(\mu)|/T}) 
  + O(e^{-A_1/T}),
\end{split}
\label{eq:eSeriesExpansion1} \\
\begin{split}
  \delta \varepsilon^{(2)}_{1,T}(\lambda)\! &=\!- \int_{-B^{(1)}}^{B^{(1)}} d \mu\, a_{1}(\lambda-\mu) \delta \varepsilon^{(1)}_{1,T}(\mu)-\int_{-B^{(1)\prime}}^{-B^{(1)}} d \mu\, a_{1}(\lambda-\mu) \varepsilon^{(1)}_{1,T}(\mu)- \int_{B^{(1)}}^{B^{(1)\prime}} d \mu\, a_{1}(\lambda-\mu) \varepsilon^{(1)}_{1,T}(\mu) 
  \\ & \quad + T \int_{-\infty}^{\infty}\! d\mu\, a_{1}(\lambda-\mu) \log (1\!+\! e^{-|\varepsilon^{(1)}_{1,T}(\mu)|/T}) \\
  & \quad + \int_{-B^{(2)}}^{B^{(2)}} d \mu\, a_{2}(\lambda-\mu) \delta \varepsilon^{(2)}_{1,T}(\mu)+\int_{-B^{(2)\prime}}^{-B^{(2)}} d \mu\, a_{2}(\lambda-\mu) \varepsilon^{(2)}_{1,T}(\mu)+ \int_{B^{(2)}}^{B^{(2)\prime}} d \mu\, a_{2}(\lambda-\mu) \varepsilon^{(2)}_{1,T}(\mu) 
  \\ & \quad - T \int_{-\infty}^{\infty}\! d\mu\, a_{2}(\lambda-\mu) \log (1\!+\! e^{-|\varepsilon^{(2)}_{1,T}(\mu)|/T}) 
  + O(e^{-A_2/T}),
\end{split}
\label{eq:eSeriesExpansion2}
\end{align}
where $B^{(r)\prime}$ is defined by
\begin{equation}
  \varepsilon^{(r)}_{1,T}(B^{(r)\prime}) = 0.
  \label{eq:pseudoFermiRapidityDefinition}
\end{equation}
Here we assume that $\varepsilon^{(r)}_{1,T}(\lambda)$ remain even and have two zeros, similarly to $\varepsilon^{(r)}_{1,0}(\lambda)$ (see Fig. \ref{fig:zeroTemperatureEpsilon}). However, the locations of the zeros are shifted from the Fermi rapidities $\pm B^{(r)}$ to $\pm B^{(r)\prime}$. 

Let us now analyse the terms in the RHS of \eqref{eq:eSeriesExpansion1}. The first term does not explicitly depend on $T$. The second and third terms will be shown later to be $O(T^4)$. Hence \eqref{eq:eSeriesExpansion1} is dominated by the fourth term, which can be expanded as
\begin{equation}
  \begin{split}
    \int_{-\infty}^{\infty}d\mu\, a_{1}(\lambda-\mu) \log (1+e^{-|\varepsilon^{(2)}(\mu)|/T}) &= \frac{T}{|\varepsilon'(B^{(2)\prime})|}  \sum_{\sigma=\pm} a_{1}(\lambda - \sigma B^{(2)\prime})  \int_{0}^{\infty} dx \log(1 + e^{-x}) + O(T^{2}) \\
    & = \frac{T\pi^{2}}{6 |\varepsilon^{\prime}(B^{(2)})|} \sum_{\sigma = \pm} a_{1}(\lambda - \sigma B^{(2)}) + O(T^{2}).
  \end{split}
  \label{eq:eFourthTermExpansion}
\end{equation}
In this expansion we used that $|B^{(2)}-B^{(2)\prime}|$ changes smoothly in the vicinity of $T=0$. After plugging \eqref{eq:eFourthTermExpansion} into \eqref{eq:eSeriesExpansion1}, we obtain 
\begin{equation}
  \delta \varepsilon^{(1)}_{1,T}(\lambda) = \frac{\pi T^{2}}{6 |\varepsilon^{(2)}_{1,T}(B^{(2)})|} \sum_{\sigma=\pm}^{} a_{1}(\lambda-\sigma B^{(2)}) - \int_{-B^{(2)}}^{B^{(2)}} d \mu\, a_{1}(\lambda-\mu) \delta \varepsilon^{(2)}_{1,T}(\mu) 
  \label{eq:eSeriesExpansionII} + o(T^{2})\,,
\end{equation}
where $o(T^{2})$ is such that 
\be
\lim_{T\rightarrow0}\frac{o(T^{2})}{T^2}=0\,.
\ee
By repeating the above procedure for $\delta \varepsilon^{(2)}_{1,T}$, we obtain
\begin{equation}
  \begin{split}
  \delta \varepsilon^{(2)}_{1,T}(\lambda) &= \frac{\pi T^{2}}{6 |\varepsilon^{(1)}_{1,T}(B^{(1)})|} \sum_{\sigma=\pm}^{} a_{1}(\lambda-\sigma B^{(1)}) - \int_{-B^{(1)}}^{B^{(1)}} d \mu\, a_{1}(\lambda-\mu) \delta \varepsilon^{(1)}_{1,T}(\mu) 
   \\
   &\quad -  \frac{\pi T^{2}}{6 |\varepsilon^{(2)}_{1,T}(B^{(2)})|} \sum_{\sigma=\pm}^{} a_{2}(\lambda-\sigma B^{(2)}) + \int_{-B^{(2)}}^{B^{(2)}} d \mu\, a_{2}(\lambda-\mu) \delta \varepsilon^{(2)}_{1,T}(\mu) + o(T^{2}),
  \label{eq:eSeriesExpansionIII}
\end{split}
\end{equation}
where we used that the second, third, sixth, and seventh terms on the RHS of \eqref{eq:eSeriesExpansion2} are $O(T^4)$. 

The equations \eqref{eq:eSeriesExpansionII}--\eqref{eq:eSeriesExpansionIII} are identical to the system \eqref{eq:FirstPseudoEnergyCorrection1}-\eqref{eq:driving4} of the main text. 

Let us now prove that the second and third terms in the r.h.s. of \eqref{eq:eSeriesExpansion1} and the second, third, sixth, and seventh terms on the RHS of \eqref{eq:eSeriesExpansion2} are $O(T^4)$. By expanding these terms for $B^{(r)\prime}\sim B^{(r)}$ and using $\varepsilon^{(r)}_{1,0}(B^{(r)})=0$ we have 
\begin{align}
&-\int_{-B^{(2)\prime}}^{-B^{(2)}} d \mu\, a_{1}(\lambda-\mu) \varepsilon^{(2)}_{1,T}(\mu)- \int_{B^{(2)}}^{B^{(2)\prime}} d \mu\, a_{1}(\lambda-\mu) \varepsilon^{(2)}_{1,T}(\mu) = O((B^{(2)\prime}-B^{(2)})^2)\,,\label{higherorder1}\\
&-\int_{-B^{(1)\prime}}^{-B^{(1)}} d \mu\, a_{1}(\lambda-\mu) \varepsilon^{(1)}_{1,T}(\mu)- \int_{B^{(1)}}^{B^{(1)\prime}} d \mu\, a_{1}(\lambda-\mu) \varepsilon^{(1)}_{1,T}(\mu)=O((B^{(1)\prime}-B^{(1)})^2)\,,\\
&\int_{-B^{(2)\prime}}^{-B^{(2)}} d \mu\, a_{2}(\lambda-\mu) \varepsilon^{(2)}_{1,T}(\mu)+ \int_{B^{(2)}}^{B^{(2)\prime}} d \mu\, a_{2}(\lambda-\mu) \varepsilon^{(2)}_{1,T}(\mu) = O((B^{(2)\prime}-B^{(2)})^2)\,.\label{higherorder3}
\end{align}
We now assume that 
\be
B^{(r)\prime}-B^{(r)}=O(T^{\alpha_r})\qquad\qquad \alpha_r>0\,.
\ee
Let us prove that $\alpha_r>0$ by \emph{reductio ad absurdum}. Suppose that $0<\alpha_2\leq1$. Then, using \eqref{eq:eSeriesExpansion1} we have 
\begin{equation}
  \delta \varepsilon^{(1)}_{1,T}(\lambda) = T^{2\alpha_2} F^{(1)}(\lambda) + O(T^{2}),
  \label{eq:reductioAdAbsurdumHypothesis}
\end{equation}
for some $F^{(1)}(\lambda)$ independent of $T$. If we then take $\delta \varepsilon^{(1)}_{1,T}(B^{(1)\prime})$ and expand around $\lambda=B^{(1)}$, we get
\begin{equation}
\delta \varepsilon^{(1)}_{1,T}(B^{(1)\prime}) = - \varepsilon^{(1)\prime}_{1,T}(B^{(1)})(B^{(1)\prime}-B^{(1)}) + O( (B^{(1)\prime}-B^{(1)})^{2} )= T^{2\alpha_2} F^{(1)}(\lambda) + O(T^{2}).
  \label{eq:reductioExpansion}
\end{equation}
So we conclude $\alpha_1=2\alpha_2$. Using now \eqref{eq:eSeriesExpansion2} we have 
\begin{equation}
  \delta \varepsilon^{(2)}_{1,T}(\lambda) = T^{2\alpha_2} F^{(2)}(\lambda) + O(T^{4\alpha_2}).
  \label{eq:reductioAdAbsurdumHypothesis2}
\end{equation}
Expanding $\delta \varepsilon^{(2)}_{1,T}(B^{(2)\prime})$ for $B^{(2)\prime}$ around $\lambda=B^{(2)}$ we then get
\be
\delta \varepsilon^{(2)}_{1,T}(B^{(2)\prime}) = - \varepsilon^{(2)\prime}_{1,T}(B^{(2)})(B^{(2)\prime}-B^{(2)}) + O( (B^{(2)\prime}-B^{(2)})^{2} )= T^{2\alpha_2} F^{(2)}(\lambda) + O(T^{4\alpha_2}).
\ee  
This is, however, not compatible with the hypothesis because it would require $\varepsilon^{(2)\prime}_{1,0}(B^{(2)})=0$. So we must have $\alpha_2>1$, implying that Eq.~\eqref{eq:eSeriesExpansionII} holds. This in turn implies that $\alpha_1>1$, indeed by expanding $\delta \varepsilon^{(1)}_{1,T}(B^{(1)\prime})$ for $B^{(1)\prime}$ around $\lambda=B^{(1)}$ we have
\begin{equation}
\delta \varepsilon^{(1)}_{1,T}(B^{(1)\prime}) = - \varepsilon^{(1)\prime}_{1,T}(B^{(1)})(B^{(1)\prime}-B^{(1)}) + O( (B^{(1)\prime}-B^{(1)})^{2} )= O(T^{2}).
  \label{eq:reductioExpansion2}
\end{equation}
Since $\alpha_r>1$ for $r=1,2$ we have that Eqs.~\eqref{eq:eSeriesExpansionII} hold \eqref{eq:eSeriesExpansionIII} hold. Expanding then $\delta \varepsilon^{(r)}_{1,T}(B^{(r)\prime})$ for $B^{(r)\prime}$ around $\lambda=B^{(r)}$ we have 
\be
B^{(r)\prime}-B^{(r)} = - \frac{\pi^{2} T^{2}}{6 (\varepsilon^{(r)\prime}_{1,0}(B^{(r)}))^{2}	} U^{(r)}(B^{(r)}) + o(T^2)
\label{eq:bChangeInB}
\ee
where the functions $U^{(r)}(x)$ are defined in the main text (\emph{cf}. Eqs.~\eqref{eq:repU1Definition}-\eqref{eq:repU2Definition}). This proves that \eqref{higherorder1}-\eqref{higherorder3} are $O(T^4)$.  

\subsection{Expansion of the particle density and velocity}
\label{sub:particleDensityVelocity}

Now we show how the low temperature corrections \eqref{eq:lowTemperatureDensitiesRDefinitions1}--\eqref{eq:lowTemperatureDensitiesRDefinitions2} and \eqref{eq:lowTemperatureVelocitiesWDefinitions1}--\eqref{eq:lowTemperatureVelocitiesWDefinitions2} are derived for the particle density $\rho^{(r)}_{n,T}(\lambda)$ and the particle velocity $v^{(r)}_{n,T}(\lambda)$. This derivation is based the expansion \eqref{eq:integral}--\eqref{eq:exp} of the integral 
\be
I_{f}^{(r)} = \int_{-\infty}^{\infty}\! \mathrm{d} \lambda\,\,\, \vartheta^{(r)}_{1,T}(\lambda) f(\lambda),
\label{eq:integralAppendix}
\ee
which is reported in \cite{BePi17}. In explicit notation the expansion reads as 
\begin{align}
  I_{f}^{(r)} &= \int_{-B^{(r)}}^{B^{(r)}} d\lambda f(\lambda)\nonumber \\ & \quad + \frac{\pi^{2} T^{2}}{6 (\varepsilon^{(r)\prime}_{1,0}(B^{(2)}))^{2}}\left[ f'(B^{(r)}) - f'(-B^{(r)}) - \left( \frac{\varepsilon^{(r)\prime\prime}_{1,0}(B^{(r)})}{\varepsilon^{(r)\prime}(B^{(r)})} + U(B^{(r)}) \right)(f(B^{(r)})+f(-B^{(r)})) \right],
  \label{eq:iIntegralExpansionFull}
\end{align}
where we neglected sub-leading contributions.

Now, considering \eqref{eq:firstrholowT}--\eqref{eq:secondrholowT} and separating the ground state root densities $\rho^{(r)}_{1,\mathrm{t},0}$ from the corrections $\delta\rho^{(r)}_{1,\mathrm{t}}$ we have  
\begin{align}
  &\rho^{(1)}_{1,\mathrm{t},0}(\lambda)+\delta\rho^{(1)}_{1,\mathrm{t}}(\lambda)=\frac{1}{2 \pi} + a_{1} \star \vartheta^{(2)}_{1,T}\rho^{(2)}_{1,\mathrm{t},0}(\lambda)+ a_{1} \star \vartheta^{(2)}_{1,T}\delta \rho^{(2)}_{1,\mathrm{t}}(\lambda),\label{eq:apprhostep11}\\
  &\rho^{(2)}_{1,\mathrm{t},0}(\lambda)+\delta\rho^{(2)}_{1,\mathrm{t}}(\lambda)=a_1 \star \vartheta^{(1)}_{1,T}\rho^{(1)}_{1,\mathrm{t},0}(\lambda)+a_1 \star \vartheta^{(1)}_{1,T}\delta\rho^{(1)}_{1,\mathrm{t}}(\lambda) - a_2\star \vartheta^{(2)}_{1,T}\rho^{(2)}_{1,\mathrm{t},0}(\lambda)- a_2\star \vartheta^{(2)}_{1,T}\delta\rho^{(2)}_{1,\mathrm{t}}(\lambda).\label{eq:apprhostep12}
\end{align}
Applying now \eqref{eq:iIntegralExpansionFull} to the convolutions and retaining corrections up to $O(T^2)$ we have 
\begin{align}
&a_{r} \star \vartheta^{(s)}_{1,T}\delta\rho^{(s)}_{1,\mathrm{t}}(\lambda)=a_{r} \ast \delta\rho^{(s)}_{1,\mathrm{t}}(\lambda)\big|_s\,, \\
&a_{r} \star \vartheta^{(s)}_{1,T}\rho^{(s)}_{1,\mathrm{t},0}(\lambda)=a_{r} \ast \rho^{(s)}_{1,\mathrm{t},0}(\lambda)\big|_s+ \frac{\pi^{2}T^{2}}{6 } a_r\ast z^{(s)}_U(\lambda)\bigr |_s\,,\qquad\qquad\qquad r,s=1,2\,, \label{eq:conv}
\end{align}
where we used the definition \eqref{eq:GSconvolution} for the ground state convolution and the function $z^{(s)}_U(\lambda)$ is of the form \eqref{eq:driving3} with the coefficients \eqref{eq:lowTemperatureDensities1}--\eqref{eq:lowTemperatureDensities3}. Plugging into \eqref{eq:apprhostep11}--\eqref{eq:apprhostep12} the expressions \eqref{eq:conv} for the convolutions and subtracting the ground state equations \eqref{eq:rZeroTemperatureDensities1}--\eqref{eq:rZeroTemperatureDensities2} we finally obtain \eqref{eq:lowTemperatureDensitiesRDefinitions1}--\eqref{eq:lowTemperatureDensitiesRDefinitions2}. 

The low-temperature expansion for the excitation velocities is obtained in a similar manner. One has to remove the exponentially suppressed bound state terms from the system \eqref{eq:vExcitationVelocitiesPhysical}--\eqref{eq:vExcitationVelocitiesAuxiliary}, obtaining
\begin{align}
  &v^{(1)}_{1,T}\rho^{(1)}_{1,\mathrm{t},T}(\lambda)=\frac{e'(\lambda)}{2 \pi}+ a_{1} \star \vartheta^{(2)}_{1,T}v^{(2)}_{1,T}\rho^{(2)}_{1,\mathrm{t},T}(\lambda), \label{eq:vExcitationVelocitiesPhysicalLowTemp} \\
  &v^{(2)}_{1,T}\rho^{(2)}_{1,\mathrm{t},T}(\lambda)=a_1 \star \vartheta^{(1)}_{1,T}v^{(1)}_{1,T}\rho^{(1)}_{1,\mathrm{t},T}(\lambda) - a_2\star \vartheta^{(2)}_{1,T}v^{(2)}_{1,T}\rho^{(2)}_{1,\mathrm{t},T}(\lambda).\label{eq:vExcitationVelocitiesAuxiliaryLowTemp}
\end{align}
Then by applying \eqref{eq:iIntegralExpansionFull} to the convolutions we find  \eqref{eq:lowTemperatureVelocitiesWDefinitions1}--\eqref{eq:lowTemperatureVelocitiesWDefinitions2}. 

\subsection{The first correction to the energy}
\label{sub:eneryHomogeneousFirstCorrection}

In this Appendix we derive the identity \eqref{eq:identity}--\eqref{eq:fFunctionDefinition2}, which is necessary to derive the low temperature correction to the energy density, $\delta e$ \eqref{eq:edenscorr}. First, it is necessary to introduce a vectorial notation for the sake of compactness. To a function $w^{(r)}(\lambda)$ we associate the vector $w$, whose components are
\begin{equation}
  [w]_{r,\lambda} = w^{(r)}(\lambda).
  \label{eq:fVectorDefinition}
\end{equation}
Operators $\hat K$ acting on these vectors are defined as
\begin{equation}
  [\hat K\, w]_{r,\lambda} = \sum_{s=1}^{2} \int_{-B^{(s)}}^{B^{(s)}} d \mu K_{rs}(\lambda-\mu) w^{(s)}(\mu).
  \label{eq:bOperatorDefinition}
\end{equation}
In this notation, the low temperature BGT equations \eqref{eq:firstrholowT}--\eqref{eq:secondrholowT} read
\begin{equation}
  \rho_{\mathrm{t}} = d_{\rho} - \hat A \hat \vartheta \rho_{\mathrm{t}},
  \label{eq:rLowTempBGTVec}
\end{equation}
where 
\begin{align}
  \rho_{\mathrm{t}} = \begin{bmatrix} \rho^{(1)}_{1,\mathrm{t},T}(\lambda) \\ \rho^{(2)}_{1,\mathrm{t},T}(\lambda)  \end{bmatrix},
  \qquad
  &[\hat \vartheta] = 
  \begin{bmatrix} 
    \vartheta^{(1)}_{1}(\lambda-\mu) & 0 \\
    0 & \vartheta^{(2)}_{1}(\lambda-\mu)
  \end{bmatrix} \cdot \delta(\lambda-\mu),
  \label{eq:rVecDefinitions}
 \\ 
 d_{\rho} = \begin{bmatrix}
    \frac{1}{2\pi} \\
    0
  \end{bmatrix},
  \qquad
  &[\hat A] = \begin{bmatrix}
    0 & -a_{1}(\lambda-\mu) \\
    -a_{1}(\lambda-\mu) & a_{2}(\lambda-\mu)
  \end{bmatrix}.
\end{align}
In the same notation, the first correction to the densities \eqref{eq:lowTemperatureDensitiesRDefinitions1}--\eqref{eq:lowTemperatureDensitiesRDefinitions2} reads 
\begin{align}
  \delta \rho_{\mathrm t,T} = \frac{\pi^{2} T^{2}}{6} R, \qquad R = d_{R} - \hat A R,
  \label{eq:rhoCorrectionVector}
\end{align}
where
\begin{align}
  \delta \rho_{\mathrm t, T} = 
  \begin{bmatrix}
    \delta \rho^{(1)}_{1,\mathrm{t},T}(\lambda) \\
    \delta \rho^{(2)}_{1,\mathrm{t},T}(\lambda)
  \end{bmatrix},
  \qquad
  d_{R} =
  - \hat A z_{R,I},
  \qquad
  z_{R,I} = 
  \begin{bmatrix}
    z^{(1)}_{R,I}(\lambda) \\
    z^{(2)}_{R,I}(\lambda)
  \end{bmatrix}.
  \label{eq:rhoCorrectionVectorDefinitions}
\end{align}
The functions $z^{(r)}_{R,I}(\lambda)$ are given by the coefficients \eqref{eq:lowTemperatureDensities1}--\eqref{eq:lowTemperatureDensities3}.

To prove the identity  \eqref{eq:identity}--\eqref{eq:fFunctionDefinition2}, we start from the left hand side of \eqref{eq:identity}
\begin{equation}
  \sum_{r=1}^{2} \int_{-\infty}^{\infty} \mathrm d \lambda (R^{(r)}(\lambda) + z^{(r)}_{R,I}(\lambda)) q^{(r)}(\lambda) .
  \label{eq:firstCorrectionEnergyDensityAppendix}
\end{equation}
Using the vector notation and \eqref{eq:rhoCorrectionVector}-- \eqref{eq:rhoCorrectionVectorDefinitions}, this correction can be written as the scalar product
\begin{equation}
  q \cdot (-(\hat 1 + \hat A)^{-1} \hat A z_{R, I} + z_{R,I}) = q \cdot (\hat 1+\hat A)^{-1} z_{R, I},
  \label{eq:firstCorrectionEnergyVector}
\end{equation}
where the scalar product is defined as
\begin{equation}
  a\cdot b = \sum_{r=1}^{2} \int_{-B^{(r)}}^{B^{(r)}}\! \mathrm d\lambda\, a^{(r)}(\lambda) b^{(r)}(\lambda). 
  \label{eq:scalarProduct}
\end{equation}
Since $\hat A$ is symmetric, we can write \eqref{eq:firstCorrectionEnergyVector} as 
\begin{equation}
  ((\hat 1 + \hat A)^{-1}q)\cdot z_{R,I} \equiv f_q \cdot z_{R,I}, \qquad f_q = q - \hat A f_q.
  \label{eq:energyExpressionVector}
\end{equation}
Writing this result in the standard TBA notation yields exactly the identity \eqref{eq:identity}--\eqref{eq:fFunctionDefinition2}.

\section{Low-temperature expansion of the density and current profiles}
\label{sec:inhomogeneousLowTemperatureExpansionAppendix}

In this Appendix we show the derivation of the low temperature expansions \eqref{eq:EvenChargeStationaryValues}--\eqref{eq:EvenCurrentStationaryValues} and \eqref{eq:EvenChargeStationaryValues2}--\eqref{eq:EvenCurrentStationaryValues2} of the charge and current density profiles after a bipartite quench. In other words, we will evaluate
\begin{align}
  q(\frak r,\zeta) &= \sum_{r=1,2} \int_{-\infty}^{\infty} \mathrm{d}\lambda \rho^{(r)}_{1,\mathrm{t},\zeta}(\lambda)  \vartheta^{(r)}_{1,\zeta}(\lambda) q^{(r)}(\lambda), 
  \label{eq:chargeDensityAppendix} \\
  j(\frak r,\zeta) &= \sum_{r=1,2} \int_{-\infty}^{\infty} \mathrm{d}\lambda \rho^{(r)}_{1,\mathrm{t},\zeta}(\lambda)  \vartheta^{(r)}_{1,\zeta}(\lambda)v^{(r)}_{1,\zeta}(\lambda) q^{(r)}(\lambda),
  \label{eq:currentDensityAppendix}
\end{align}
at low temperatures. In \eqref{eq:chargeDensityAppendix}--\eqref{eq:currentDensityAppendix} we have already used the fact that bound states are exponentially suppressed at small enough temperatures. 

The basis of the evaluation of \eqref{eq:chargeDensityAppendix}--\eqref{eq:currentDensityAppendix} is the expansion \eqref{eq:lFarFromLightconeCondition}--\eqref{eq:LowTemperatureIntegralSeries2}, carried out in detail in \cite{BePi17}. This expansion has two cases according to whether the ray $\zeta$ is close to one of the light cones $\pm v^{(r)}_{1,0}(B^{(r)})$ or not. We treat these two cases separately. Our computations below are valid up to $O(T^{2})$.

\subsection{The case $\lim_{T\rightarrow 0} |\zeta-v^{(r)}_{1,0}(B^{(r)})| \ne 0$}
\label{sub:farFromLightCone}
Far away from the light cones, one can use the low temperature expansion \eqref{eq:LowTemperatureIntegralSeries1}. First we compute the correction to the charge density
\begin{equation}
  \delta q(\frak r,\zeta) = \sum_{r=1,2} \int_{-\infty}^{\infty} \mathrm{d}\lambda \rho^{(r)}_{1,\mathrm{t},\zeta}(\lambda)\vartheta^{(r)}_{1,\zeta}(\lambda) q^{(r)}(\lambda) - \sum_{r=1,2} \int_{-B^{(r)}}^{B^{(r)}} \mathrm{d}\lambda \rho^{(r)}_{1,\mathrm{t},0}(\lambda) q^{(r)}(\lambda). 
  \label{eq:qVectorInh}
\end{equation}
The vectorial form of the $O(T^{2})$ expansion \eqref{eq:LowTemperatureIntegralSeries1} is
\begin{equation}
  I_{\zeta,f} = f \cdot j + \frac{\pi T^{2}}{6} f \cdot z_{I,\zeta},
  \label{eq:expansionVecInh}
\end{equation}
where
\begin{equation}
  I_{\zeta,f} = 
  \begin{bmatrix}
    I^{(1)}_{\zeta,f} \\
    I^{(2)}_{\zeta,f}
  \end{bmatrix}, 
  \quad
  f = 
  \begin{bmatrix}
    f^{(1)} \\
    f^{(2)}
  \end{bmatrix},
  \quad
  j = 
  \begin{bmatrix}
    1 \\
    1
  \end{bmatrix},
  \quad
  z_{I,\zeta} = 
  \begin{bmatrix}
    z_{I,\zeta}^{(r)} \\
    z_{I,\zeta}^{(r)}
  \end{bmatrix}.
  \label{expansionVecInh}
\end{equation}
Using expansion \eqref{eq:expansionVecInh} on the BGT equations \eqref{eq:firstrholowT}--\eqref{eq:secondrholowT} yields
\begin{align}
  \delta \rho_{\mathrm t,\zeta} = \frac{\pi^{2} T^{2}}{6} R_{\zeta}, \qquad R_{\zeta} = d_{R,\zeta} - \hat A R_{\zeta},
  \label{eq:rhoCorrectionVectorInh}
\end{align}
where
\begin{align}
  \delta \rho_{\mathrm t, \zeta} = 
  \begin{bmatrix}
    \delta \rho^{(1)}_{1,\mathrm{t},\zeta}(\lambda) \\
    \delta \rho^{(2)}_{1,\mathrm{t},\zeta}(\lambda)
  \end{bmatrix},
  \qquad
  d_{R,\zeta} =
  - \hat A z_{R,I,\zeta},
  \qquad
  z_{R,I,\zeta} = 
  \begin{bmatrix}
    z^{(1)}_{R,I,\zeta}(\lambda) \\
    z^{(2)}_{R,I,\zeta}(\lambda)
  \end{bmatrix},
  \label{eq:rhoCorrectionVectorDefinitionsInh}
\end{align}
and the symbol $\delta$ denotes difference from the ground state value, i.e.,
\begin{equation}
  \delta \rho_{1,\mathrm{t},\zeta} = \rho_{1,\mathrm{t},\zeta}-\rho_{1,\mathrm{t},0}.
  \label{eq:deltaMeanign}
\end{equation}
The function $z^{(r)}_{R,I,\zeta}(\lambda)$ is
\be
\begin{split}
z_{R,I,\zeta}^{(r)}(\lambda)&= z_{R,I}^{(r)}(\lambda) \Theta_H [-v^{(r)}_{1,0}(B^{(r)})-\zeta]+( \frak r^2 z_{R,I,-}^{(r)}(\lambda)+ z_{R,I,+}^{(r)}(\lambda))\,  \Theta_H [v^{(r)}_{1,0}(B^{(r)})-|\zeta|] \\ & \quad + \frak r^2 z_{R,I}^{(r)}(\lambda)\Theta_H [\zeta-v^{(r)}_{1,0}(B^{(r)})],
\end{split}
   \label{eq:zRFunction}
\ee
with $z_{R,I}^{(r)}(\lambda)$ and $z_{R,I,\pm}^{(r)}(\lambda)$ given by the coefficients \eqref{eq:lowTemperatureDensities1}--\eqref{eq:lowTemperatureDensities3}.

 Plugging \eqref{eq:rhoCorrectionVectorInh} in \eqref{eq:qVectorInh} and using the expansion \eqref{eq:expansionVecInh}, we get 
\begin{equation}
  \delta q(\frak r, \zeta) = \frac{\pi^{2} T^{2}}{6} q \cdot (-(\hat 1+\hat A)^{-1} \hat A z_{R,I,\zeta} + z_{R,I,\zeta} ) = \frac{\pi^{2} T^{2}}{6} q \cdot ((\hat 1+\hat A)^{-1} z_{R,I,\zeta}).
  \label{eq:firstCorrectionEnergyVectorInh}
\end{equation}
Since $\hat A$ is symmetric, we can write \eqref{eq:firstCorrectionEnergyVectorInh} as
\begin{equation}
  \delta q(\frak r, \zeta) = \frac{\pi^{2} T^{2}}{6} ((\hat 1 + \hat A)^{-1}q)\cdot z_{R,I,\zeta} \equiv \frac{\pi^{2} T^{2}}{6} f_q \cdot z_{R,I,\zeta}, \qquad f_q = q - \hat A f_q.
  \label{eq:energyExpressionVectorInh}
\end{equation}
This is exactly \eqref{eq:EvenChargeStationaryValues} in vectorial notation, which is valid up to $O(T^{2})$. We note that the above expansion is analogous to the homogeneous case shown in Appendix \ref{sub:eneryHomogeneousFirstCorrection}. The only difference is the dependence of the vector $z_{R,I,\zeta}$ on $\zeta$.

In the case of the charge current density $\delta j(\frak r,\zeta)$, the logic of the derivation is the same. We have
\begin{equation}
  \delta j(\frak r,\zeta) = \sum_{r=1,2} \int_{-\infty}^{\infty} \mathrm{d}\lambda \rho^{(r)}_{1,\mathrm{t},\zeta}(\lambda)  \vartheta^{(r)}_{1,\zeta}(\lambda)v^{(r)}_{1,\zeta}(\lambda) q^{(r)}(\lambda) - \sum_{r=1,2} \int_{-\infty}^{\infty} \mathrm{d}\lambda \rho^{(r)}_{1,\mathrm{t},0}(\lambda)  v^{(r)}_{1,\zeta}(\lambda) q^{(r)}(\lambda).
  \label{eq:jVectorInh}
\end{equation}
In vectorial notation, the equations \eqref{eq:vExcitationVelocitiesPhysicalLowTemp}--\eqref{eq:vExcitationVelocitiesAuxiliaryLowTemp} for the velocities become
\begin{equation}
  w_{\zeta} = d_{w} - \hat A \hat \vartheta_{\zeta} w_{\zeta},
  \label{eq:wEquation}
\end{equation}
with
\begin{equation}
  w_{\zeta} = 
  \begin{bmatrix}
    \rho^{(1)}_{1,\mathrm{t},\zeta}(\lambda) v^{(1)}_{1,\zeta}(\lambda) \\
    \rho^{(2)}_{1,\mathrm{t},\zeta}(\lambda) v^{(2)}_{1,\zeta}(\lambda) 
  \end{bmatrix},
  \quad
  d_{w} = 
  \begin{bmatrix}
    \lambda / (2\pi) \\
    0
  \end{bmatrix}.
  \label{eq:wEquationSource} 
  \end{equation}
Using the expansion \eqref{eq:expansionVecInh} on \eqref{eq:wEquation} yields 
\begin{equation}
  \delta w_{\zeta} = \frac{\pi T^{2}}{6} W_\zeta,\quad W_\zeta = d_{W,\zeta} - \hat A W_\zeta,
  \label{eq:wLowTempInh}
\end{equation}
where
\begin{equation}
  d_{W,\zeta} = \hat A z_{W,I,\zeta}, \quad z_{W,I,\zeta} = 
  \begin{bmatrix}
    z^{(1)}_{W,I,\zeta}(\lambda) \\
    z^{(2)}_{W,I,\zeta}(\lambda)
  \end{bmatrix}.
  \label{eq:zwDefinitionsInh}
\end{equation}
The functions $z^{(r)}_{W,I,\zeta}(\lambda)$ are given by
\begin{equation}
  \begin{split}
    z_{W,I,\zeta}^{(r)}(\lambda)&= z_{W,I}^{(r)}(\lambda) \Theta_H [-v^{(r)}_{1,0}(B^{(r)})-\zeta]+( \frak r^2 z_{W,I,-}^{(r)}(\lambda)+ z_{W,I,+}^{(r)}(\lambda))\,  \Theta_H [v^{(r)}_{1,0}(B^{(r)})-|\zeta|]+ \\ &\qquad \frak r^2 z_{W,I}^{(r)}(\lambda)\Theta_H [\zeta-v^{(r)}_{1,0}(B^{(r)})],
   \label{eq:zWFunction}
 \end{split}
 \end{equation}
with $z_{W,I}^{(r)}(\lambda)$ and $z_{W,I,\pm}^{(r)}(\lambda)$ given by the coefficients \eqref{eq:lowTemperatureWCD1}--\eqref{eq:lowTemperatureWCD3}.

Plugging \eqref{eq:wLowTempInh} into \eqref{eq:jVectorInh} and using the expansion \eqref{eq:expansionVecInh}, we get
\begin{equation}
  \delta j(\frak r, \zeta) = \frac{\pi^{2} T^{2}}{6} q \cdot (-(\hat 1+\hat A)^{-1} \hat A z_{W,I,\zeta} + z_{W,I,\zeta} ) = \frac{\pi^{2} T^{2}}{6} q \cdot ((\hat 1+\hat A)^{-1} z_{W,I,\zeta}).
  \label{eq:firstCorrectionCurrentVectorInh}
\end{equation}
Since $\hat A$ is symmetric, we can write \eqref{eq:firstCorrectionEnergyVectorInh} as
\begin{equation}
  \delta j(\frak r, \zeta) = \frac{\pi T^{2}}{6} ((\hat 1 + \hat A)^{-1}q)\cdot z_{W,I,\zeta} \equiv \frac{\pi^{2}T^{2}}{6}f_q \cdot z_{W,I,\zeta}, \qquad f_q = q - \hat A f_q.
  \label{eq:currentExpressionVectorInh}
\end{equation}
This is exactly \eqref{eq:EvenCurrentStationaryValues} in vectorial notation, which is valid up to $O(T^{2})$.

\subsection{The case $\zeta\pm v^{(r)}_{1,0}(B^{(r)})\sim O(T)$}
\label{sub:closeToLightCone}
Close to the light cones, we use the expansion \eqref{eq:LowTemperatureIntegralSeries2} instead of \eqref{eq:LowTemperatureIntegralSeries1} but otherwise the procedure is the same as in the far-from-lightcone case of Appendix \ref{sub:farFromLightCone}. We can get the final results by replacing $z_{R,I,\zeta}$ and $z_{W,I,\zeta}$ with $z_{II,\zeta}$ in Eqs. \eqref{eq:energyExpressionVectorInh} and \eqref{eq:currentExpressionVectorInh}. The final result is
\begin{align}
  \delta q(\frak r, \zeta) &=\frac{\pi^2 T(1-\frak r^2)}{6} ((\hat 1 + \hat A)^{-1}q)\cdot z_{II,\zeta} \equiv \frac{\pi^2 T(1-\frak r^2)}{6} f_q \cdot z_{II,\zeta}, \qquad f_q = q - \hat A f_q,
  \label{eq:energyExpressionVectorInhLightCone} \\
  \delta j(\frak r, \zeta) &= \frac{\pi^2 T(1-\frak r^2)}{6} ((\hat 1 + \hat A)^{-1}q)\cdot z_{II,\zeta} \equiv \frac{\pi^2 T(1-\frak r^2)}{6} f_q \cdot z_{II,\zeta}, \qquad f_q = q - \hat A f_q.
  \label{eq:currentExpressionVectorInhCone}
\end{align}
These are exactly \eqref{eq:EvenChargeStationaryValues2}--\eqref{eq:EvenCurrentStationaryValues2} in vectorial notation, valid up to $O(T)$.

\end{document}